# Contents

# Analysis and Modeling of Traffic in Modern Data Communication Networks
Ohio State University
Department of Computer and Information Science

G. Babic, B. Vandalore, R. Jain

February 5, 1998

**Abstract**

In performance analysis and design of communication netword modeling data traffic is important. With introduction of new applications, the characteristics of the data traffic changes. We present a brief review the different models of data traffic and how they have evolved. We present results of data traffic analysis and simulated traffic, which demonstrates that the packet train model fits the traffic at source destination level and long-memory (self-similar) model fits the traffic at the aggregate level.

# 1 Introduction

Characteristics of data traffic play crucial role in performance analysis and design of communication networks. Understanding the models of computer network traffic will help us design better protocols, design better network topologies, design better routing and switching hardware and provide better services to the users.

We believe that:

- a sudden exponential up turn in the last several years of number of hosts interconnected in networks (e.g. in Internet since 1988 [JAI95]) and intensified usage of network application, which in turn increased amount of traffic in network several order of magnitude,

and

- introduction and intensive usage of new network applications, which have appeared recently, such as WWW, Gopher and newsgroups (quite different from traditional network applications), have completely changed nature and characteristics of network traffic, from those traffic had been exercised until mid 1980's.

Just as a complement to the previous statement and as justification of necessity for correct understanding of traffic characteristics, we provide a citation from [PAR93]: "... we still do not understand how data communications traffic behaves. After nearly a quarter century of data communications, researchers are still struggling to develop accurate traffic models. Yet daily we must make decisions about how to configure networks and build networking devices based on our inadequate models of data traffic. There's a serious need for more research work on non-Poisson queuing models."

In summary, exceptional growth in amount of network traffic and emerging of new types of traffic clearly indicate that new studies and research in the area are needed. The above reasons are the motivation for this research.

In our research, we performed collection of data traffic from an academic computer network, then we concentrate on traffic analysis and modeling, primarily for purpose of discovering influence of traffic characteristics on network design parameters.

This report is organized as follows. Section 2, provides overview of different data traffic models, starting with Poisson (classical) model, which is followed two more recent models: packet train model and long-memory (self-similar) model. In Section 3, we give description of our tests and analysis (using long-memory and packet train models) of data traffic we obtained from CIS (Computer and Information Sciences Department) network. At the end we provide an extensive bibliography.

# 2 Overview of Traffic Modeling

In this Section we give a brief overview three traffic models.

Stochastic models of packet traffic used in past were almost exclusively Markovian in nature, or more generally short-range dependent traffic processes. Those traffic models, now called classic models, assumed Poisson arrival rate and exponential length of messages. Data source models with those characteristics were used in the analysis and modeling of early ARPANET and agreement between real data and results from queuing models were good and satisfactory. And so were agreements in 10-15 years that followed.

It is worth noting that traditional telephony has beneficed very much (for understanding its internal behavior and for a system design) for a long time (more than a half century) from classic traffic models, which basically assume that call holding times are exponentially distributed. But, some recent studies [DUF94] have shown that call holding times may be best described using heavy-tailed distributions with possibly infinite variance/mean. These characteristics are contrary to those of exponential distribution. The very possible reason for the change in characteristics of telephone traffic is that telephony systems are now being used not only for its traditional voice communication but more and more for computer communication (e.g. remote access), telex traffic and other new telematic services. Apparently those new traffics have quite different characteristics than voice communication and as their share of overall traffic has grown, so overall traffic characteristics have changed from classical ones.

Similarly, in mid 80's, it became apparent that traditional traffic models for computer networks were less appropriate, because predicted performances and real data would not agree, and many practitioners noticed that discrepancy. Again, probably nothing was wrong (although some are trying to establish exactly that) with early queuing models of computer network traffic and they were appropriate for their times, but nature of the traffic has changed since, as result of different usage of computer networks.

Many studies indicate considerable increase in overall amount of traffic in the network. For example, [STI95] mentions the Internet's 20 percent average monthly increases on its most heavily used segment. That by itself may qualitatively change the nature of the traffic. But, noticeable are also types of traffic generated by new network applications such as World Wide Web, Gopher and newsgroups, which are quite different from traditional application such as file transfer protocol (FTP), remote access (telnet) and E-mail (smtp). These new types of traffic can obviously change overall traffic characteristics in computer networks.

As result of those observations and trends, since mid 1980's, research in the area of traffic characterization and its implications on design of computer networks has intensified [JAI86, JAI90, LEL91, FOW91, GAG94a, LEL94a, KLI94a, ADA95, PAX94a, PAX94b]. The concept of "packet trains" was introduced in 1986 [JAI86] and become very popular. This model assumes that a group of packets travel together as a train, contrary to the Poisson model which assumes that packets are independent, so it can be seen as "car model".

One of features of today's networking traffic discovered 3-4 years ago is its long-range dependency (long-memory) [COX84], found in both local area networks [LEL94a] and wide-area networks [KLI94a]. In addition, traffic in local area network is self-similar (fractal) [LEL94a], characteristics usual not found in wide-area traffic [KLI94b], which appears to be asymptotically self-similar in only few recorded traces.

Now we provide some details of each of mentioned traffic models.

## 2.1 Poisson (Classical) Model

First studies on data traffic [JAC69, FUC70] indicated that the data traffic sources in communication networks were often bursty in nature, i.e. relatively short sequence of source activities are followed by long idle periods. During 70's and early 80's, those and some other studies suggested the following assumptions as reasonable, if somewhat simplified, for external data sources:

1. The interarrival times of messages generated by an external data source are exponentially distributed, i.e., each external data source behaves as a Poisson process. Let $G(i), i = 1, 2, \ldots, N$, be a random variable denoting interarrival times of messages generated from the ith data source.

2. The length of messages generated by an external data source are exponentially distributed. Let $H(i), i = 1, 2, \ldots, N$ be a random variable denoting length of messages from the ith data source.

3. Processes described by random variables $G(i)$ and $H(i)$ are stationary and independent.

As a consequence of the assumption 1, the aggregate traffic from several data sources would get smoother and smoother with an increase of a number of sources.

The assumption 2, about exponential distribution of lengths of messages can be relaxed to general distribution, and still closed-form solutions for different statistical parameters (mean, variance and other moments of a distribution) could be obtained using queuing theory methods.

Assumption 3, above is not realistic, because at least interarrival times and message lengths for message streams entering a communication switch (node) are clearly statistically dependent. For example, if we consider two successive messages arriving into a switch, the second message cannot get into the switch before the first message has arrived completely.

Kleinrock [KLE64] observed that problem in the exact mathematical analysis of store-and-forward communication networks and resolved it by introducing the *independence assumption.* That assumption basically states that each time message is received at a switch a new length is chosen for this message from an exponential distribution. It implies that as the message length is changing from node to node, the service time (transmission time at each link) is not identical for the same message as it passes through the network. Clearly this assumption does not correspond to the actual situation, but its mathematical consequences (applicability of the Jackson's theorem [JAC63]) resulted in models which accurately described the behavior of store-and-forward communication networks.

## 2.2 Packet Train Model

The packet train model assume that a group of packets travel together, and it should be obvious that a protocol design based on the assumption of a train arrival would be quite different from one based on independent arrivals. In the car model, each car has to decide at each intersection (or exit) whether to take exit or not. Even if all packets are going to one destination, they each make independent decision, which may result in unnecessary overhead. The overhead is apparent on computer networks in which all intermediate nodes (routers, gateways, or bridges) must make this decision for all packets. In a train model, on the other hand, the locomotive (the first packet of the train) may make the routing decision, and all other packets of a train may follow it.

It must be pointed out that the packet train model is a source model. It applies only when we look at the packets coming or going to a single node. Unlike the Poisson processes, trains are not additive. The sum of a number of trains is not a train.

In order to allow analytical modeling with a simplified form of train model, usage of a two-state Markov model is suggested. The source can be either in generation (train) state or idle (inter-train) state. The transitions between these states are memoryless (Markovian). The duration of the two states is exponentially distributed, with intertrain arrival times usually of the order of several seconds and intercar times inside the trains of the order of a few milliseconds.

## 2.3 Long-Memory (Self-Similar) Model

Some recent studies show that packet traffic in modern networks is strongly auto-correlated and there exists a long-range dependency, i.e. a persistence in their correlation structures does not die even for large lags.

### 2.3.1 Theoretical Background

For a stochastic process $X = (X_t : t = 0, 1, 2, \ldots)$ to be a *second order* (*weak or covariance or wide-sense*) stationary, it is sufficient to have the existence of a stationary mean $\mu = E[X_t]$, a stationary and finite variance $v = E[(X_t - \mu)^2]$, and a stationary auto-covariance (function) associated with a process of observations made at successive times $k = cov(X_t, X_{t+k}) = E[(X_t - \mu)(X_{t+k} - \mu)], k = 0, 1, 2 \ldots$, that depends only on $k$ and not on $t$.

Let $X = (X_t : t = 0, 1, 2, \ldots)$ be a second order stationary stochastic process, with a mean $\mu = E[X_t]$, a variance $v = E[(X_t - \mu)^2]$, and auto-covariance (function) $\gamma_k = cov(X_t, X_{t+k}) = E[(X_t - \mu)(X_{t+k} - \mu)], k = 0, 1, 2, \ldots$; Note, $v = \gamma_0$. Let the auto-correlation (function) of $X$ at lag $k$ be denoted as $\rho_k$, and by definition $\rho_k = \gamma_k / \gamma_0$.

We can think of a packet traffic process $X$ consisting of a set $\{X_t\}$, where $X_t$ is the number of packets that arrive in the t-th time unit.

(Note: We consider only stationary processes, so anywhere in this text when we use the term "auto-covariance (function)", we could use the term "auto-correlation (function) instead".)

Let let $X^{(m)} = (X_j^{(m)} : j = 1, 2, 3, \ldots)$ for each $m = 1, 2, 3, \ldots$, be the new second order stationary process, obtained by averaging the original process X over non-overlapping blocks of size m, i.e. $X_j^{(m)} = (X_{jm-m+1} + X_{jm-m+2} + \ldots + X_{jm})/m$, with the variance $v_m$, the auto-covariance $\gamma_k^{(m)}$ and the auto-correlation $\rho_k^{(m)}$. It can be shown that:

$$v_m = v/m + 2/m^2 \sum_{k=1}^{m} (m-k)\gamma_k \tag{1}$$

or

$$v_m = v/m + 2/m^2 \sum_{s=1}^{m-1} \sum_{k=1}^{s} \gamma_k \tag{2}$$

The stochastic process $X$ is said to have short-range dependency if $\sum_k \gamma_k$ is convergent, (i.e. $\sum_k \gamma_k < \infty$). Equivalently, from equation (2)

$$v_m \approx v'/m \qquad \text{with } v' \text{ finite, for large } m \tag{3}$$

('$\approx$' means that expressions on the two sides are asymptotically proportional to each other)

Such process is also called a stationary process with short memory or short-range correlations or weak dependence. An example of a stationary short-range dependent process would be a stochastic process with exponentially decaying auto-covariance function, i.e.

$$\gamma_k \approx r' a^k \qquad \text{for large } k, 0 < a < 1$$

(Note: At least in theory, other forms are possible as long as $\sum_k \gamma_k < \infty$)

Assuming 3 holds, it can be shown that $\gamma_k^{(m)} \to 0$, for $k = 1, 2, \ldots$, for large m. Then, it can be concluded that the aggregated (averaged) processes $X^{(m)}$, derived from the short-range dependent process $X$, for large $m$ tend to covariance (second order) stationary white (pure) noise.

A stochastic process $X$ is said to have *long range dependency* if $\sum_k \gamma_k$ is divergent, i.e. $\sum_k \gamma_k \to \infty$. Equivalently, from 1

$$mv_m \to \infty, \qquad \text{for large } \ m$$

Such process is also called a stationary process with long memory or long-range correlations or strong dependence. An example of a stationary long-range dependent process would be a stochastic process with hyperbolically decaying auto-covariance function, i.e.

$$\gamma_k \approx r'k^{-\beta} \qquad \text{for large } \ k, 0 < \beta < 1$$

or equivalently

$$v_m \approx v'm^{-\beta} \qquad \text{for large } \ m, 0 < \beta < 1$$

Relationship between $r'$ and $v'$ can be found from equation 1.

(Note: At least in theory, other forms are possible as long as $\sum_k \gamma_k \to \infty$) From an intuitive point, possibly the most enlightening property is that the averaged process $X^{(m)}$ takes a nondegenerated correlation structure for large m. An implication is that the averaged process $X^{(m)}$ will not appear as white noise. Instead, the (typical) aggregated traffic will have bursty subperiods and less bursty subperiods for small as well large time-scales.
It can be shown that for long-range dependent processes

$$\rho_k^{(m)} \to \rho_k \qquad \text{for large } k \text{ and } m \tag{4}$$

Equation 4 indicates that for k and m large enough, auto-correlation does not depend on m, but only on k. This property is called *asymptotic second-order self-similarity*. So, long-range dependency implies asymptotic second-order self-similarity.

The process $X$ is said to be *exactly second-order self-similar* (or *fractal*) if

$$\rho_k^{(m)} = \rho_k \qquad \text{for all } \ m, k >= 0$$

and

$$v_m = vm^{-\beta} \qquad \text{for all } \ m, k >= 0$$

Above implies that the process X and the averaged processes $X^{(m)}$ have identical correlational structure and “look” alike.

Usually, instead of the parameter $\beta$, the parameter $H = 1 - \beta/2$ is used and it is called *Hurst* coefficient. H (Hurst) coefficient characterizes the stochastic processed as follows:

1. for $1/2 < H < 1$, the process has long-range dependence,
2. for $H <= 1/2$, this is the case of a process with short-range dependence or independence.

### 2.3.2 Estimation of H Coefficient

There are several methods which can be used to estimate H coefficient [BER94], including:

- R/S statistics,
- log-log correlogram,
- log-log plot of $v_m$ versus m,
- semivariogram,
- least squares regression in the spectral domain.

All of them are heuristic approaches, each with certain problems.In our analysis for long-memory we used approach with log-log plot of $v_m$ versus m.

For large values of m, the points in the log-log plot of $v_m$ versus m are expected to be scattered around a straight line with negative slope $\beta(= 2H-1)$. In the case of a process is with short-range dependence or independence we would have $\beta = -1$ $(H = 1/2)$. For a process with long-range dependence the slope is less steeper but still negative, i.e. $-1 < \beta < 0$, implying $1/2 < H < 1$. The case that the points in a graph are scattered around a straight line with $-1 < \beta < 0$ for all values of m would imply self-similar property. It should be noted that slight departures from $H = 1/2$ are not usually easy to distinguish from the case $H = 1/2$. Similar problems appears with other methods.

### 2.3.3 Experimental Queuing Analysis

Unexpected degree of cell loss in the first ATM switches has been reported, e.g. [CSE94]. It is believed that buffer requirements was underestimated in design of those ATM switches, as consequence of modeling under traditional (Poisson) assumptions. And it is exactly that point: the implications of a long-memory traffic are quite different from ones of a Poisson traffic.

The study in [ERR94a, ERR94b, ERR94c] illustrates very nicely that:

1. a real Ethernet traffic they collected in an research and development network has quite different characteristics from a traffic generated by Markovian processes,

2. long-range dependency is an important characteristics of that real traffic,

and in this Section we shall describe their findings.

They considered the queuing system with the following characteristics:

- infinite waiting room,

- deterministic service times,

- single server and

- arrivals taken (in different ways) from actual Ethernet traffic traces.

Different transformations of original traffic trace are made, in order to preserve the marginal interarrival time distribution throughout the different queuing experiments. Service time is changed to obtain different value pairs for utilization and the average waiting times for a given input trace.

In the average delay vs. utilization plot, they first considered the following curves (see the graph 1):

- the curve A is obtained with original traffic traces;

- the curve C is obtained with an input trace that is obtained by shuffling the time series of interarrival times of original traffic trace. By randomizing the set of interarrival times, we preserve the marginal distribution of interarrival times, while destroying all correlations between them;

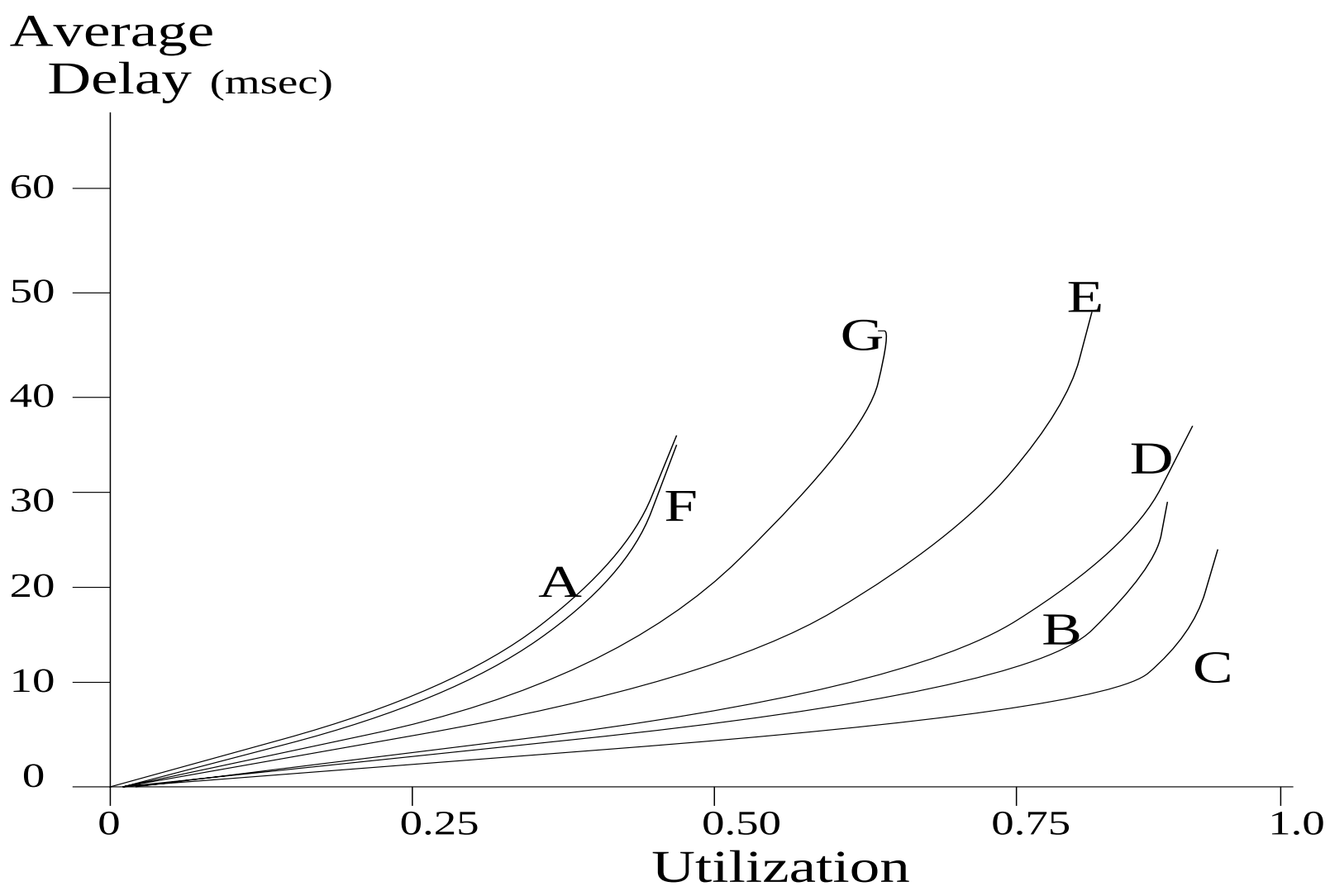

Figure 1: Average delay of transformations

- the curve B is obtained using GI/G/1 approximations based on two moment characterizations of original traffic, that are widely applied in practice.

It is noticeable that there is significant disparity between curves A and C. The curve A has a sharp rise in the average delay around 50%, while the knee of the curve C is around 90%. The implication is that the curve C is predicting much better performances than they are with real traffic.

It is interesting to note the relationship between the curve B and curve C. The curve B predicts little more conservative performance than curve C, with the knee at little over 85%. It would be an excellent result and a design guideline, if curve C was an indication of real performance. But, the curve C would indicate real performance, only if the original traffic traces were uncorrelated and independent (shuffling for the curve C destroy any such characteristics). Once more, stochastic models of packet traffic used in past are almost exclusively Markovian in nature, or, more generally short-range dependent traffic processes, and it was generally accepted that relationship between analytical results and real performances is somewhat similar to the relationship between curve B and curve C.

Several more curves are then introduced:

- the curve D is obtained with an input trace that is obtained using a

special procedure, from the original time series of interarrival times, preserving only one-step correlation and destroying all other correlations between them;

- the curve E is obtained with an input trace that is obtained after external shuffle of the original trace with m=25. External shuffle is done as follows. First, the original sequence of interarrival times is divided into blocks of size m. With N interarrival times, there are N/m such blocks. Then, the order of blocks is shuffled, while preserving the sequence inside each block. For choices of m in the range 10-100, the effect is of preserving the short range correlations while eliminating the long range correlations;

- the curve G is obtained with an input trace that is obtained after external shuffle with m=500;

- the curve F is obtained with an input trace that is obtained after internal shuffle with m=25. Internal shuffle is done as follows. First, the original sequence of interarrival times is divided into blocks of size m. With N inter-arrival times, there are N/m such blocks. Then, while preserving the sequence of blocks, a sequence of interarrival times inside each block is randomized. This has effect of destroying the short-range correlations in data, while preserving the long-range correlations.

The following can be observed from the graph:

1. The curve D (one with preserving only itemstep correlation and destroying all other correlations in the trace) is very close to the curve B (GI/G/1 approximation), indicating that something else but one-step correlation is influencing performance.

2. The curve E (external shuffle with m=25, preserving the short range correlations while eliminating the long range correlations) is still far away from the curve A. Similar results are obtained over the range of m, indicating that such models do not capture significant aspects of the queuing performance.

3. The curve F (internal shuffle with m=25, destroying the item range correlations in data, while preserving the itemrange dependency) fits almost perfectly the curve A. This indicates the long-range dependency is dominant characteristics for queuing performance, implying the same (dominant) effect on many traffic engineering design issues.

4. The curve G (external shuffle with m=500) is significantly off the curve A, indicating that correlations over extremely long time scales have measurable and practical consequence.

In conclusion, this experiment has shown that dominant characteristics for queuing performances of Ethernet traffic is a long range dependency, while the influence of a itemrange dependency is relatively small and even negligible.

# 3 Initial OSU Results in Data Traffic Analysis and Modeling

This research team has at its disposal as a testbed, for data collection and for other necessary types of research experiments, a computer network of Department of Computer and Information Science (CIS department) at the Ohio State University. This is one of largest networks of instructional, distributed, diskless work-stations in the country. Based on Ethernet technology, the computer network includes over 650 work-stations (HP715/64, Sun SLC and Sun ELC), over 100 micros (Apple Macintoshes) and about 60 file servers and minis (HP 725/64, HP 9000/755, Sun SPARC 10 and Sun 4/75).

Analysis of Ethernet traffic characteristics of network environment at the Bellcore Morris Research and Engineering Center, as a typical research or software development environment, has been reported in [LEL94a]. We believe that analysis of current Ethernet traffic in typical university environment, as CIS department, may provide different insight into traffic characteristics of modern computer networks.

We have found exceptionally intensive traffic in our network during academic quarters. For example, even on one Ethernet segment we considered moderately loaded (which includes 6 file servers, 15 dial in access ports and 20 work-stations) outside the "busy hour" (morning around 9:00am in winter quarter; students and professors start with their intensive work later), it took less than 40 minutes to collect 1,000,000 packets. On the other hand, during periods between quarters, when students leave the campus, traffic intensity is much lower. We expect that those variations in traffic intensity, with consequent changes in its characteristics, could provide additional inside in traffic behavior of university network environment and computer networks in general.

As data collecting tool we use the *tcpdump* program, distributed with Unix operating system. *Tcpdump* runs as the only active program on a dedicated data collection computer (currently Sun SPARC 10). We found that timestamps are accurate to within 15 microsec, which is appropriate from a statistical point of view. Also, the packet loss is less than 0.05%, which is again appropriate and should not have any influence to the results of statistical analysis. Monitoring computer has high capacity local disk, so monitoring and collecting of packets can be performed for a long time without any influence on the system under investigation.

Although we used some existing statistical software packages for some of required statistical analysis, in the most of cases we needed to develop new software.

Now we present results of our traffic analysis. In our analysis we used one data set obtained from monitoring real data traffic (and which includes 1 million of packets) as well as several data sets generated by simulation. Packet train and long-memory (self-similar) models were used. For long-memory (self-similar) modeling to estimate H-coefficient value we used log-log plot of vm versus m.

In the following discussion we refer to graphs shown at the end of this paper. Each graph indicate number of basic time intervals (Deltas), average number of arrivals in each time interval, with variation and standard deviation, value for H-coefficient, its confidence interval and $R^2$ (sum of the distances of the points from the best fit line).

1. For our 1 million data set, using time unit of 10 msec, H-parameter=0.853, which is similar to Leland's results, indicating self-similar nature of our Ethernet traffic. Also, it appears that the graph indicates asymptotic self-similarity. This test obviously indicates that there are dependencies in our data set. (see Graph 1)

2. For data set obtained from Uniform distribution, H-parameter=0.579, which is quite close to 0.5 (theoretical value for H parameter of uniform distribution is 0.5), indicating that our statistical programs are correct. (see Graph 2)

3. For data set obtained from Poisson distribution, H-parameter=0.543,

which is quite close to 0.5 (theoretical value for H-parameter of Poisson distribution is 0.5), indicating that our statistical programs are correct. (see Graph 3)

4. For data set obtained from our original 1 million data set by shuffling (randomizing) it, H-parameter=0.546. By randomizing the set of interfamily times, we preserve the marginal distribution of interarrival times, while destroying all correlations (independencies) between them. In the case of uncorrelated (independent) data, the value for H-parameter is close to 0.5 is correct. This test is an additional indication that our statistical programs are correct. This test does not indicate what types of dependencies exist in the data set. (see Graph 4)

5. H-parameter=0.629 is found with an input trace that is obtained from our original 1 million data set after external shuffle of the original trace with $mm = 25$. External shuffle is done as follows. First, the original sequence of interarrival times is divided into blocks of size mm. With N=1 million interarrival times, there are $N/mm$ such blocks. Then, the order of blocks is shuffled, while preserving the sequence inside each block. The effect of this external shuffle is of preserving the short range correlations while eliminating the long range correlations. Value for H-parameter (still close to 0.5) indicates that the data set has more significant long-range dependency component than that of short-range dependency. (see Graph 5)

6. H-parameter=0.732 is found with an input trace that is obtained from our original 1 million data set after external shuffle of the original trace with $mm = 100$. Conclusion from point 5. above is further supported with this test, because H-parameter increased its value, indicating that we preserved more of long-range dependencies, but even beyond mm=100 there is still significant component of long-range dependency (For the original data set H-parameter=0.853). (see Graph 6)

7. For data set obtained from packet train model (with average intertrain interarrival time=250, exponentially distributed; average number of cars in train=50, geometrically distributed; average intercar interarrival time=25, exponentially distributed), H-parameter=0.514. Note that in this case we model traffic only between two points (i.e. a railway is used with trains between two stations). H-parameter value is as expected, i.e. no dependency. (see Graph 7)

8. For data set obtained from packet train model as in point 7. but different parameters (with average intertrain interarrival time = 2500, exponentially distributed; average number of cars in train = 50, geometrically distributed; average intercar time=50, exponentially distributed), H-parameter=0.613. Note in this case values for parameters are taken to correspond to one pair of commutating nodes from the original data set. (see Graph 8)

9. Data set is obtained from packet train model, assuming existence of 10 pairs of communicating nodes over the same railway (traffic parameters for all 10 pairs were identical to those at point 7). H-parameter=0.919. (see Graph 9)

   Also, interesting insight is provided by the Graphs 9a-9h, which provide values of *H-parameter* $2, 3, 4, \ldots, 9$ pairs, respectively, of communicating nodes over the same railway (traffic parameters for all pairs were identical to those at point 7). It can be observed that H value increases with a number of pairs. This is very an important result, which states that with number of pairs traffic is becoming more and more long-range dependent, although traffic of each individual pair has no dependency.

10. Data set is obtained from packet train model, assuming existence of 10 pairs of communicating nodes over the same railway (traffic parameters for 10 pairs were identical to those of 10 most active pairs from the original set, i.e. each pair had different characteristics). With H-parameter=0.964, we again have long-range dependent traffic which is composed of traffic which have no dependency. (See Graph 10)

    From points 9 and 10, we may conclude that we have means to generate traffic which has long memory and which is self-similar. Once more, note that we made this result merging data obtained from packet model.

11. Self-similar model analysis of traffic between pair of nodes obtained filtering of the original data set is given on the Graphs 11a-11j.

# 4 Conclusion

The network traffic characteristics change as new types of applications are introduced. A traffic model can be become outdated because of the above reason. In this report we have presented an overview of three traffic models. We have done analysis of monitored and simulated traffic and shown that the packet train model and self-similar model for Ethernet traffic are not contradictory. At source and destination (between two nodes) the traffic fits the packet train model and the aggregate traffic fits the self-similar model.

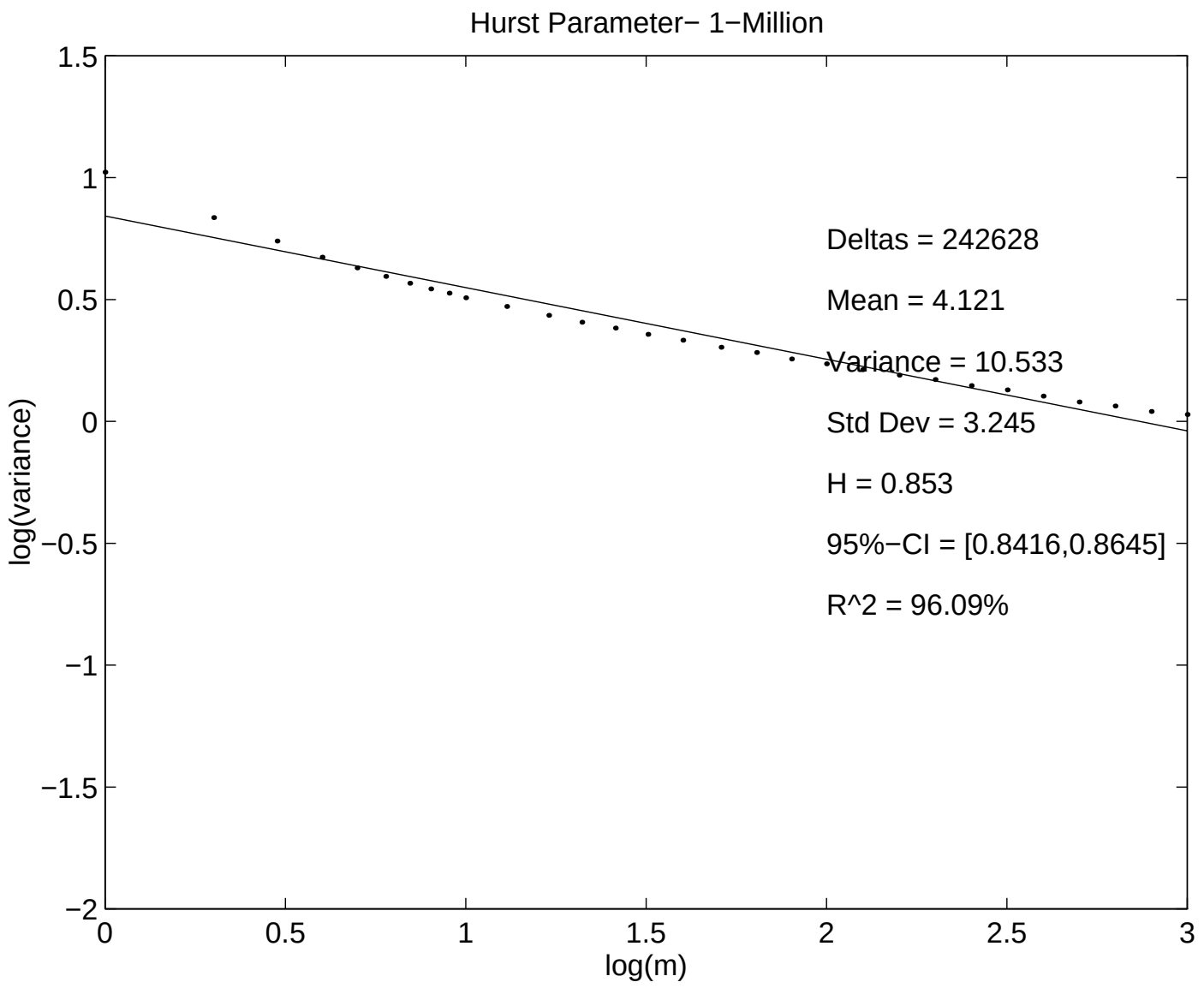

(a) 1 Million Packets

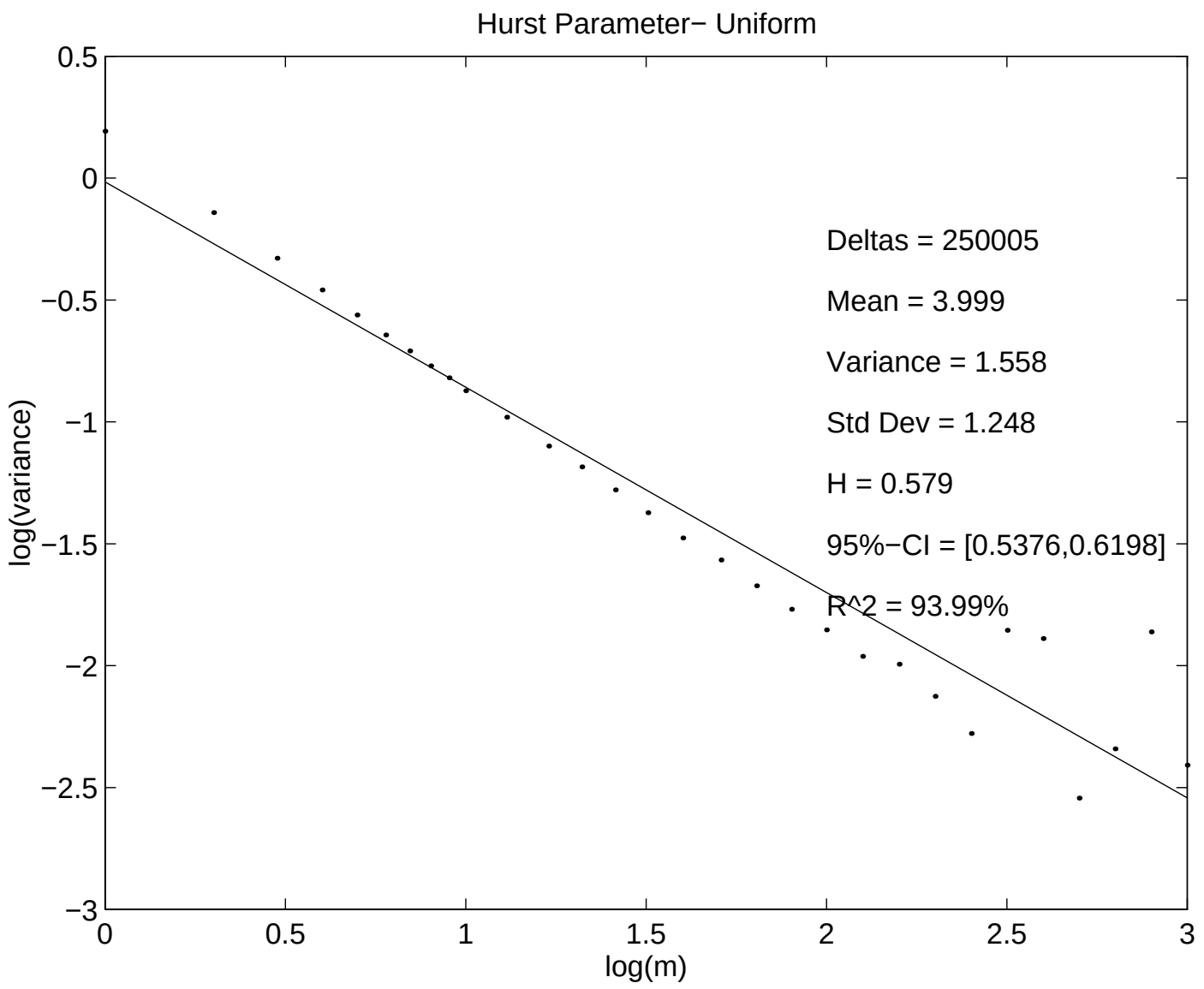

(b) Uniform Distribution

Figure 2: Graph 1 & 2

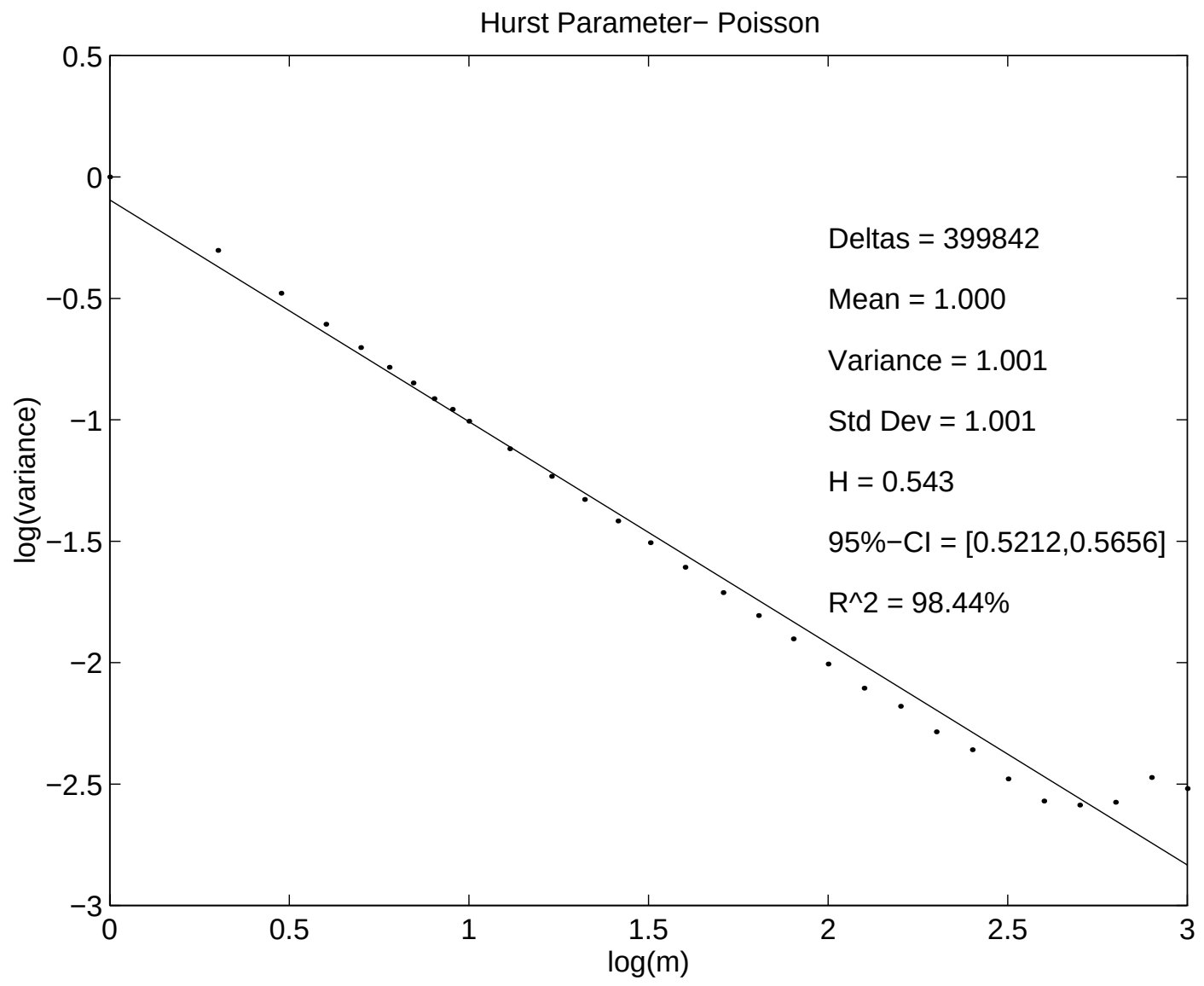

(a) Poisson Distribution

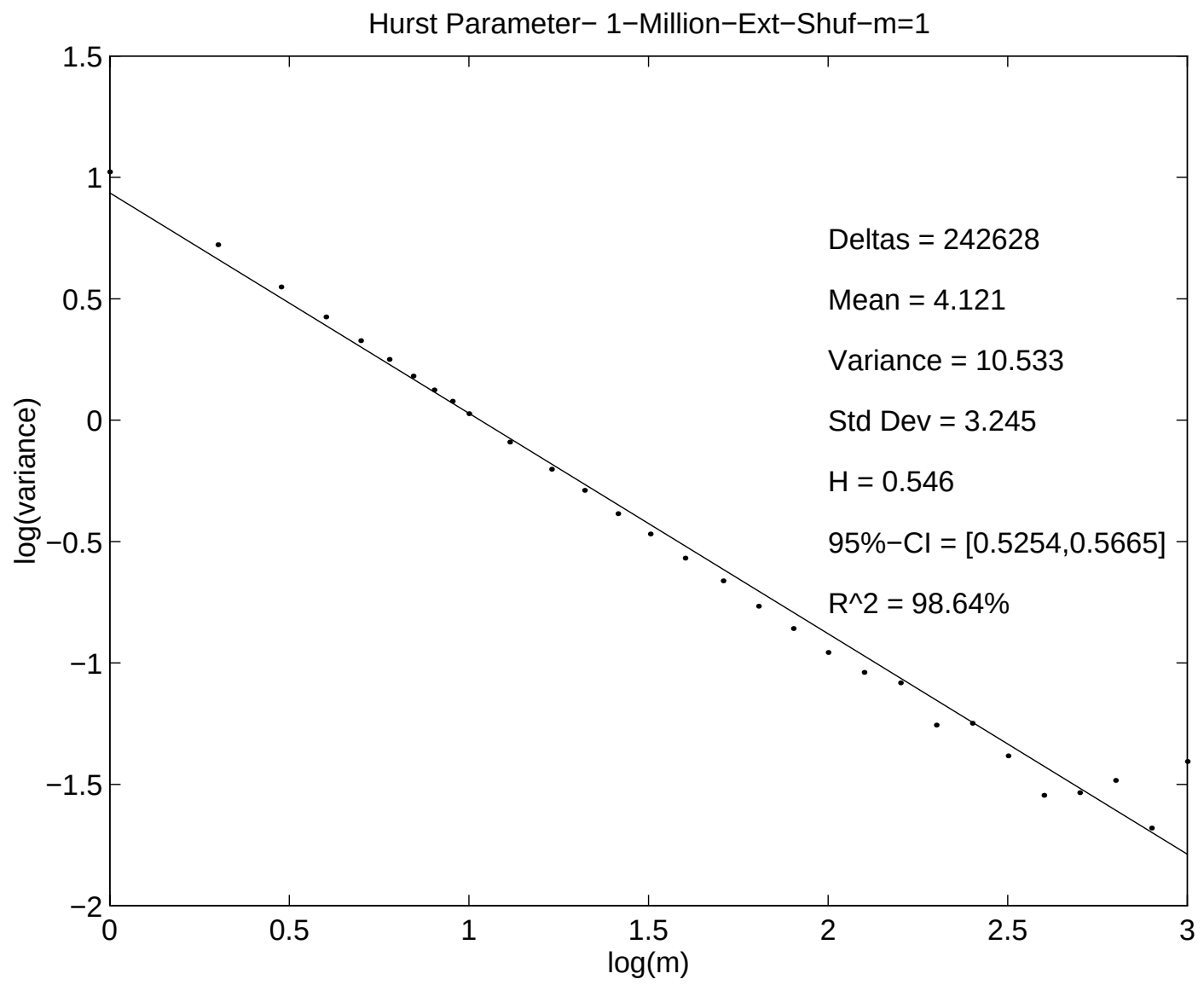

(b) External Shuffled, m=1

Figure 3: Graph 3 & 4

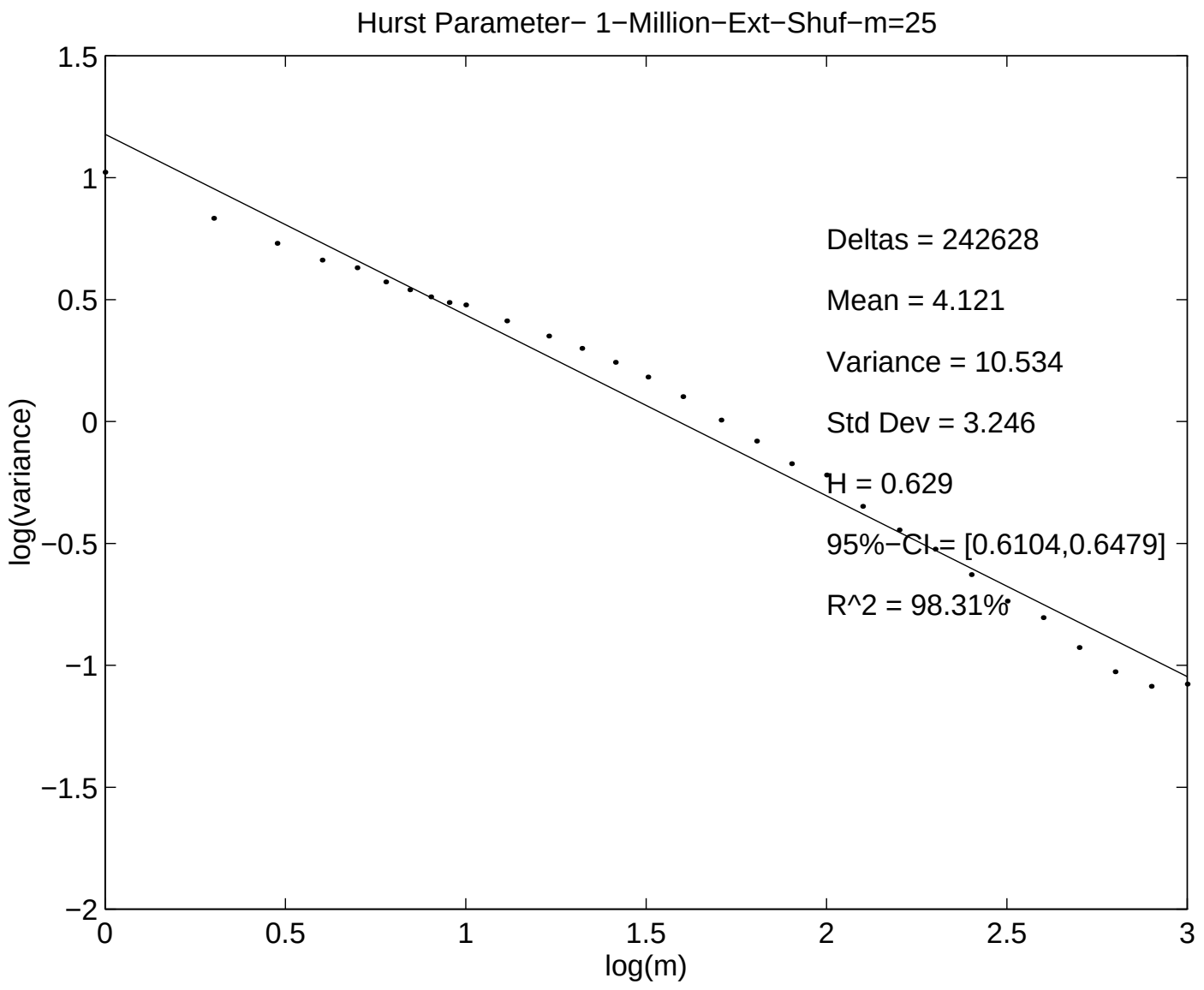

(a) External Shuffled, m=25

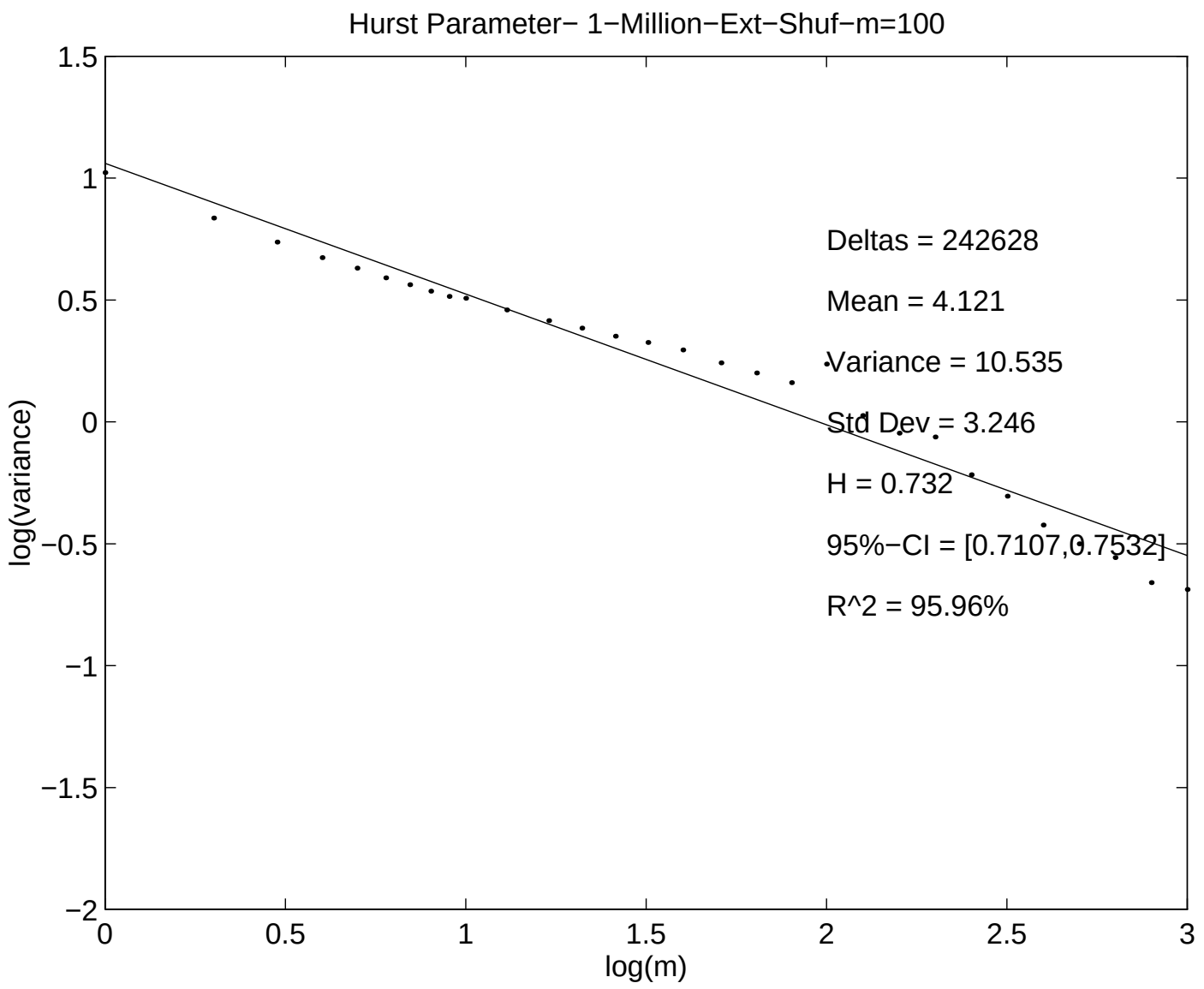

(b) External Shuffled, m=100

Figure 4: Graph 5 & 6

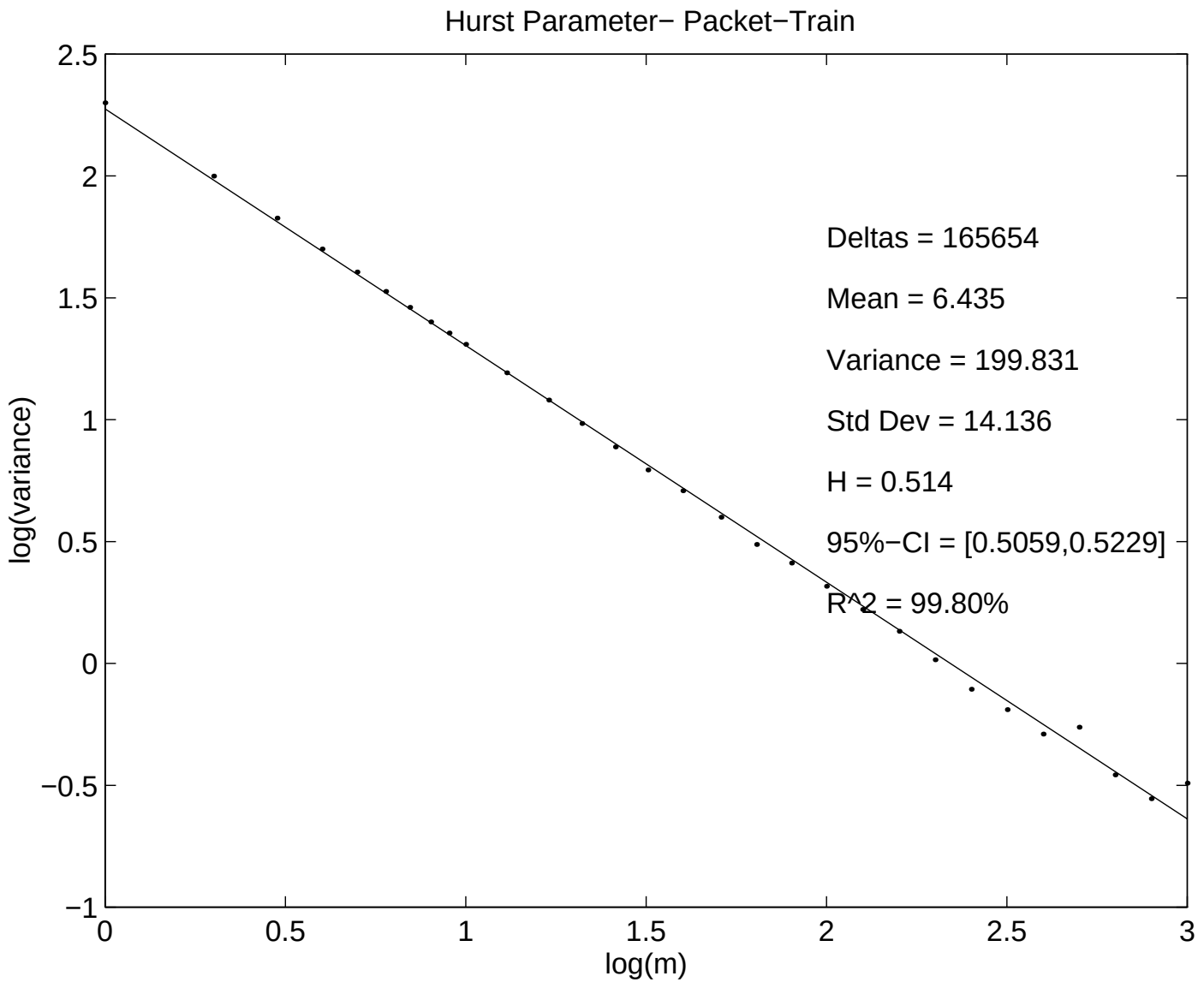

(a) Generated Packet Train 1

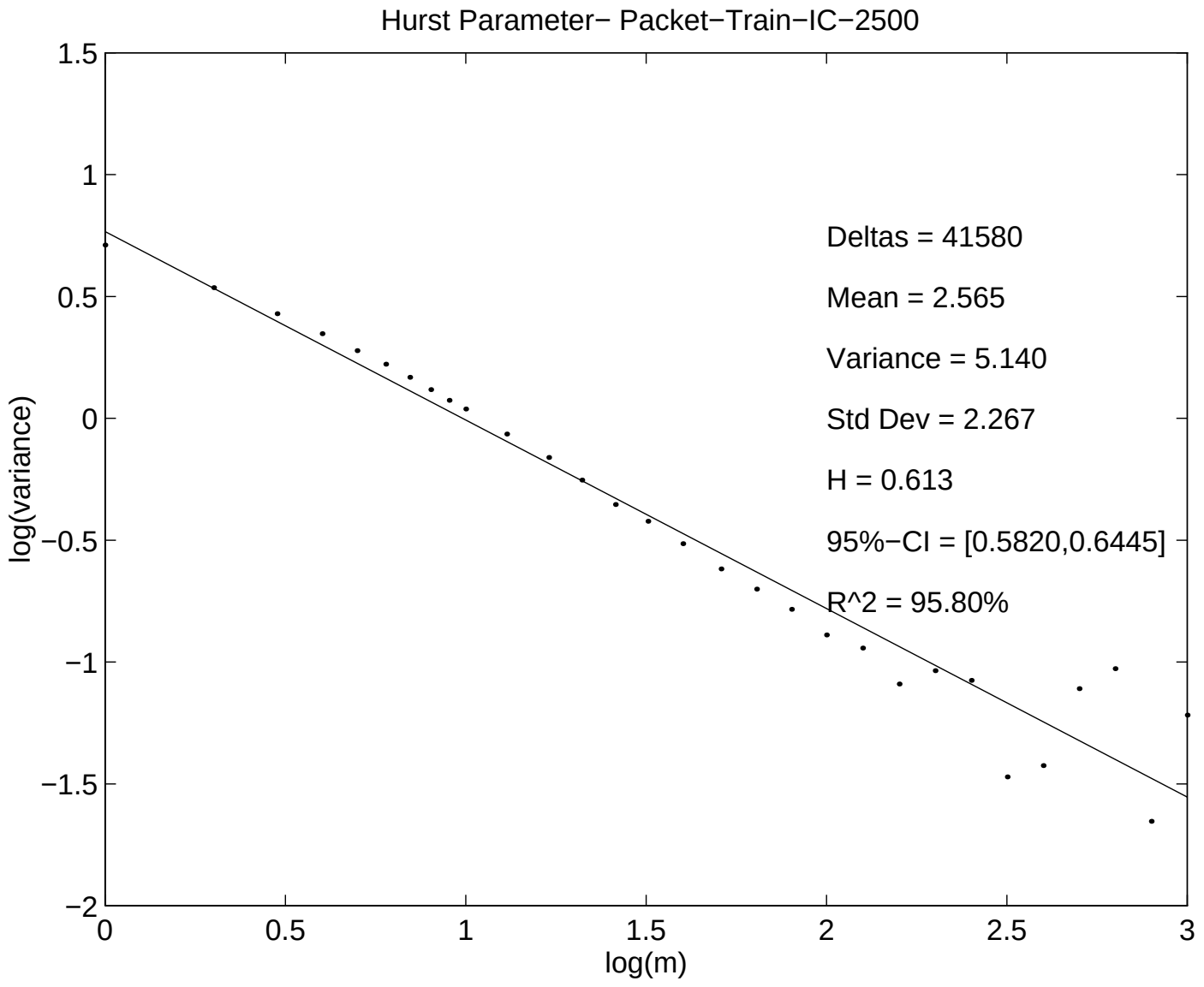

(b) Generated Packet Train 2

Figure 5: Graph 7 & 8

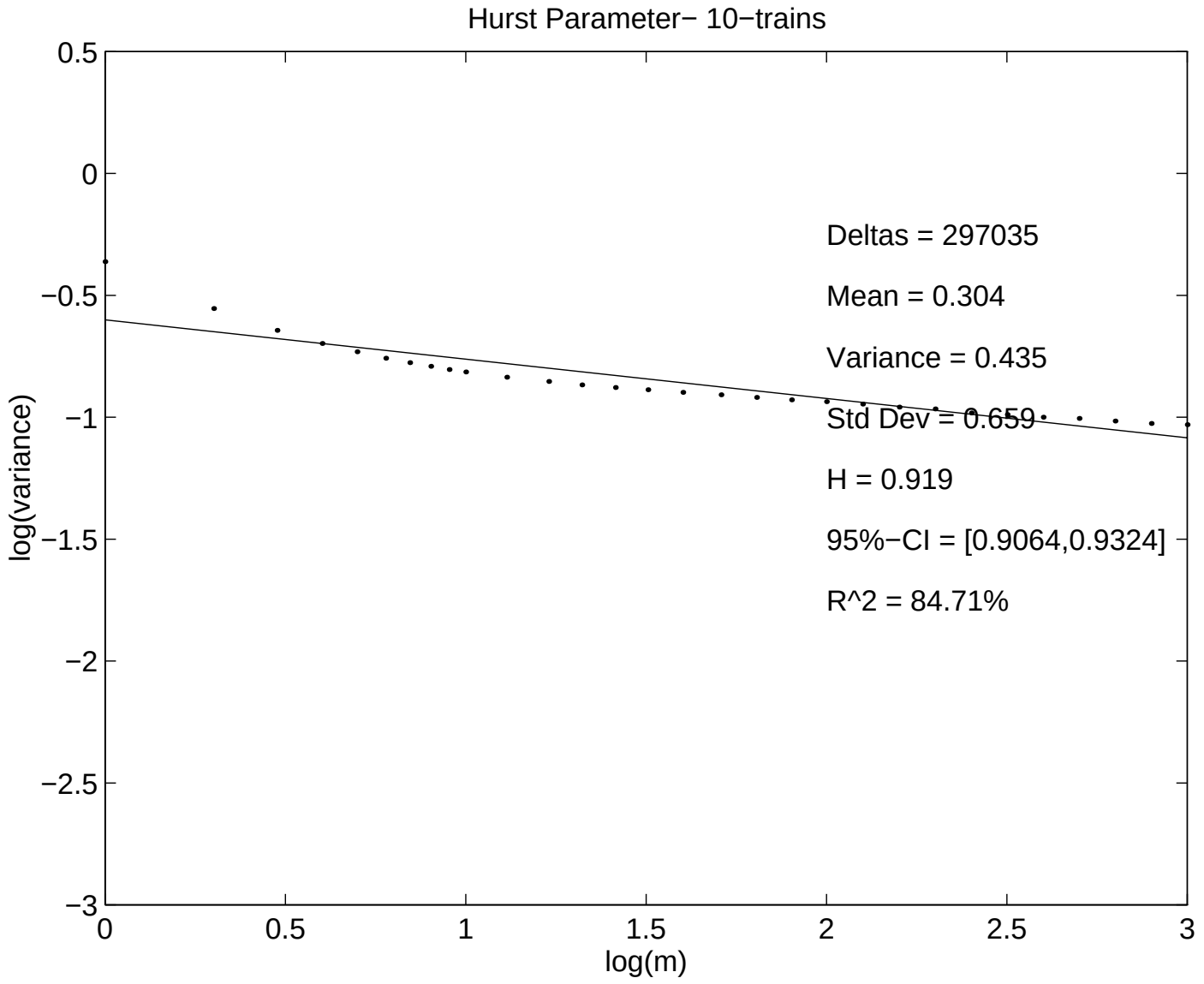

(a) 10 Generated Packet Trains

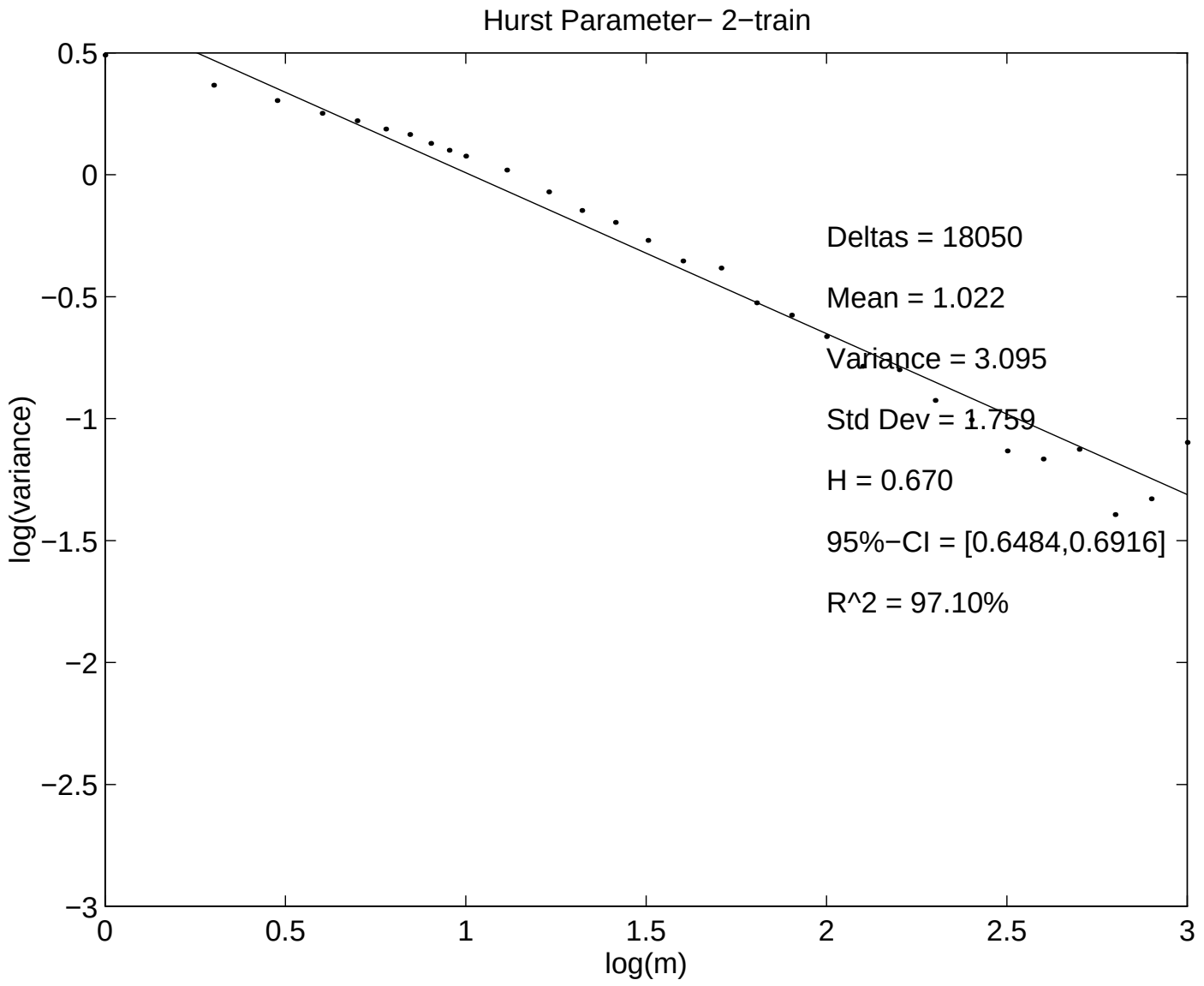

(b) 2 Generated Packet Trains

Figure 6: Graph 9 & 9a

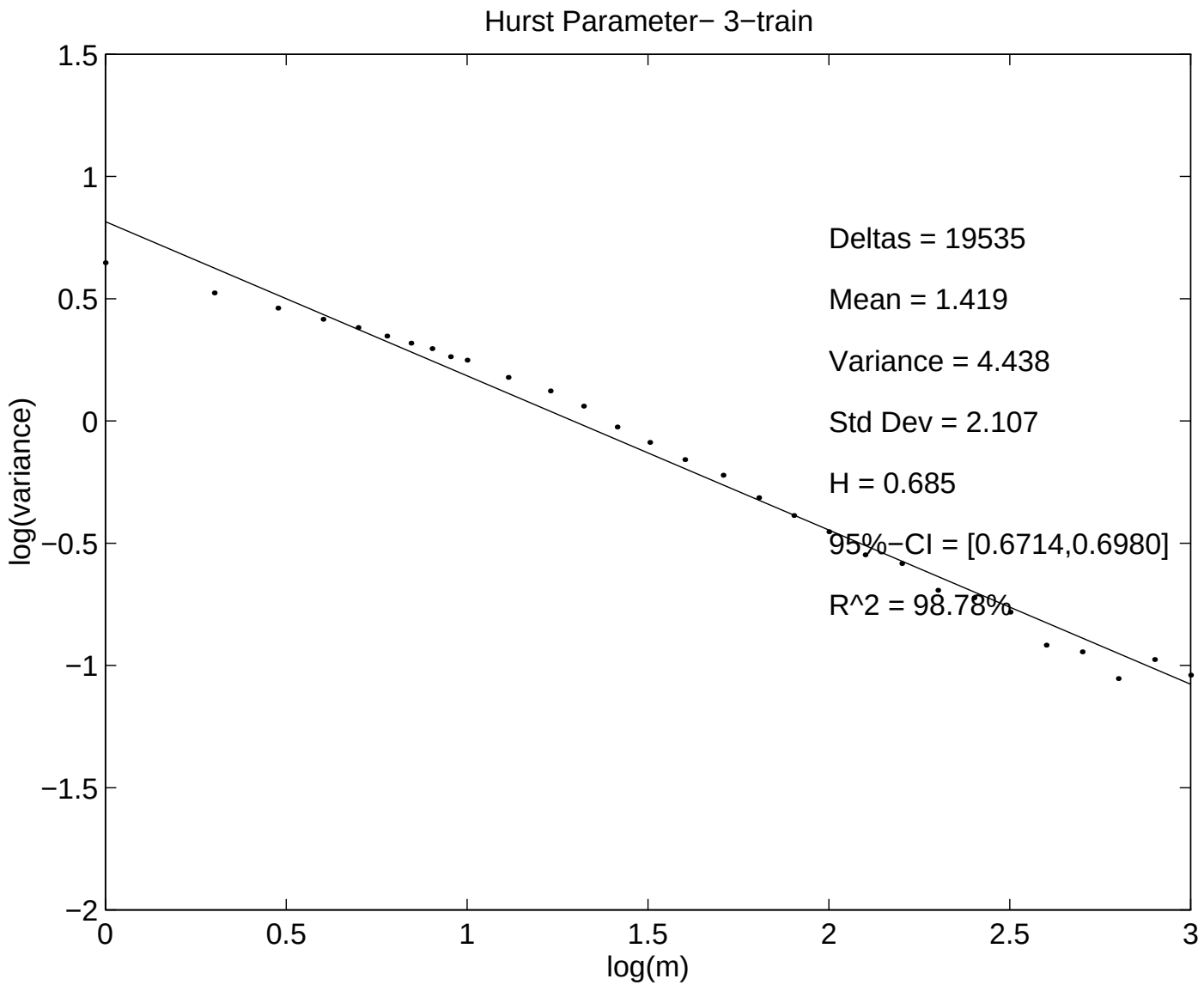

(a) 3 Generated Packet Trains

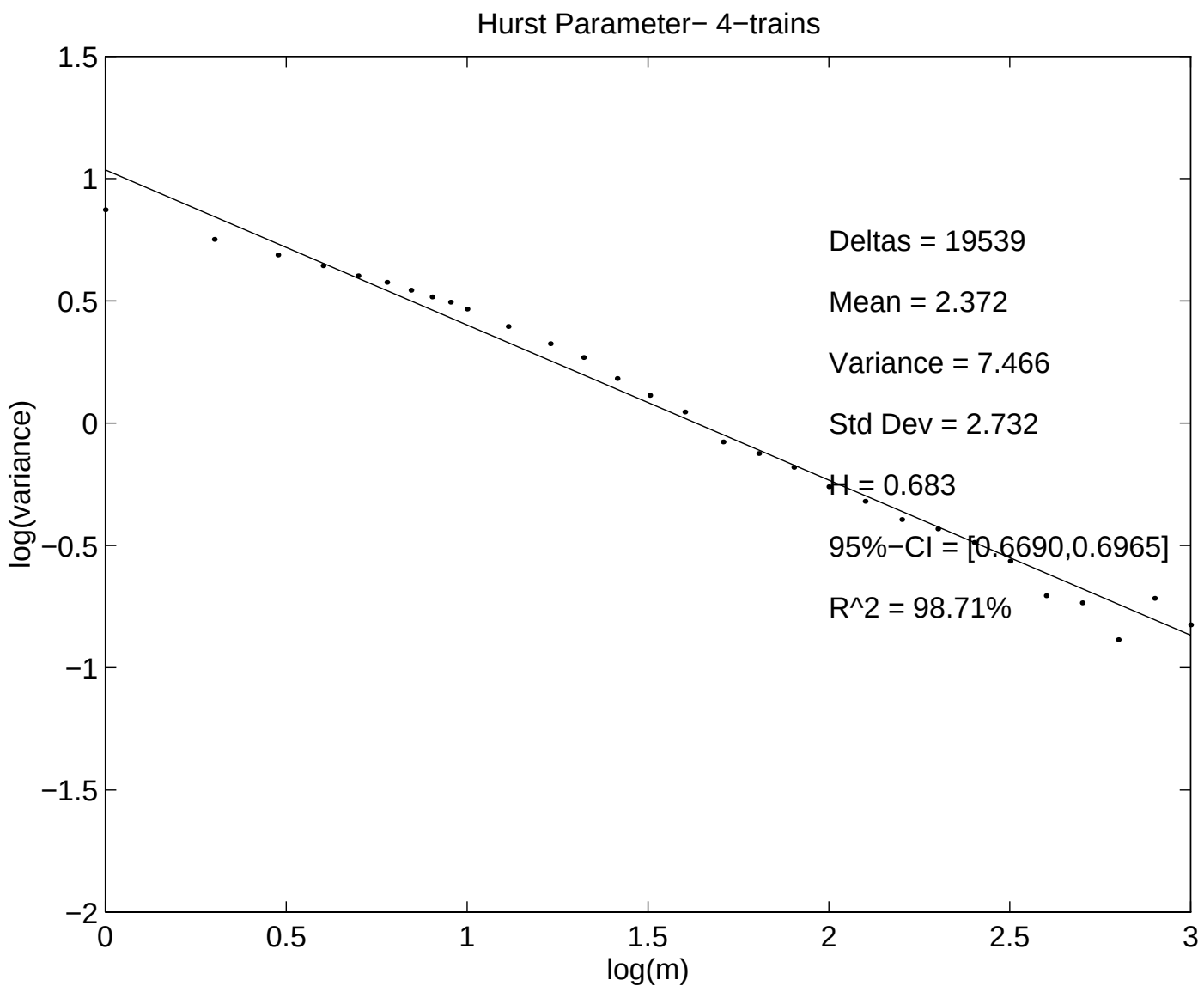

(b) 4 Generated Packet Trains

Figure 7: Graph 9b & 9c

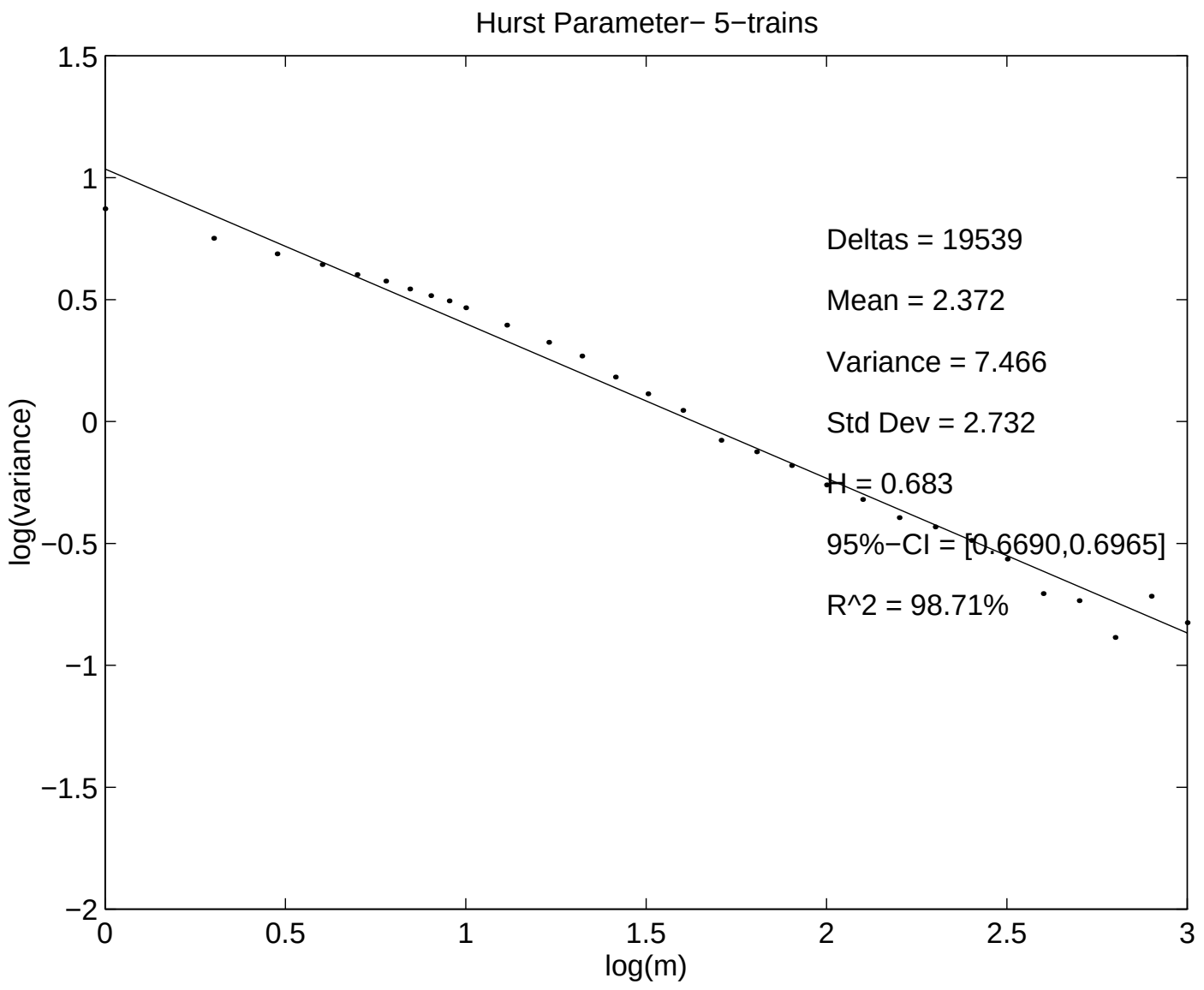

(a) 5 Generated Packet Trains

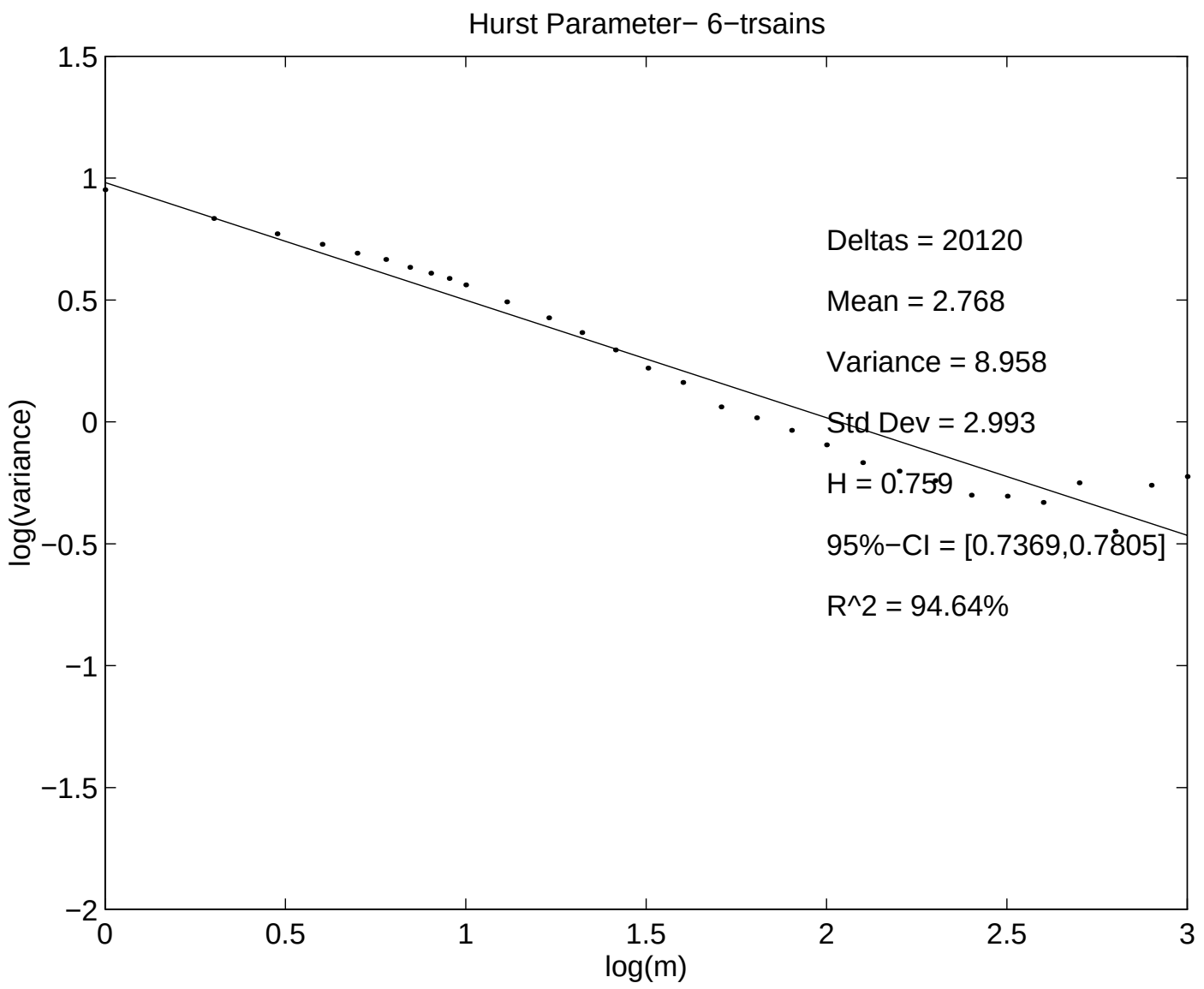

(b) 6 Generated Packet Trains

Figure 8: Graph 9d & 9e

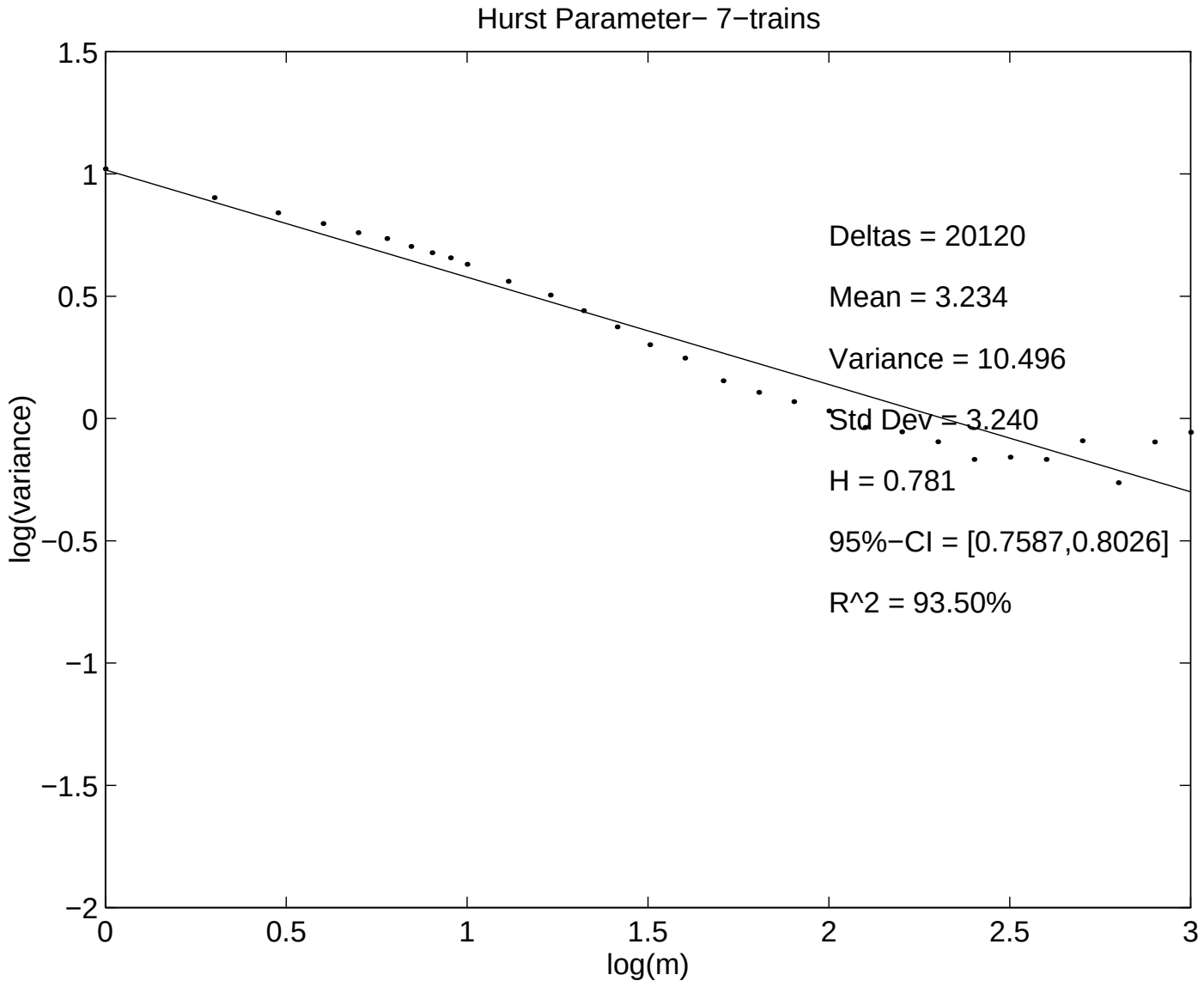

(a) 7 Generated Packet Trains

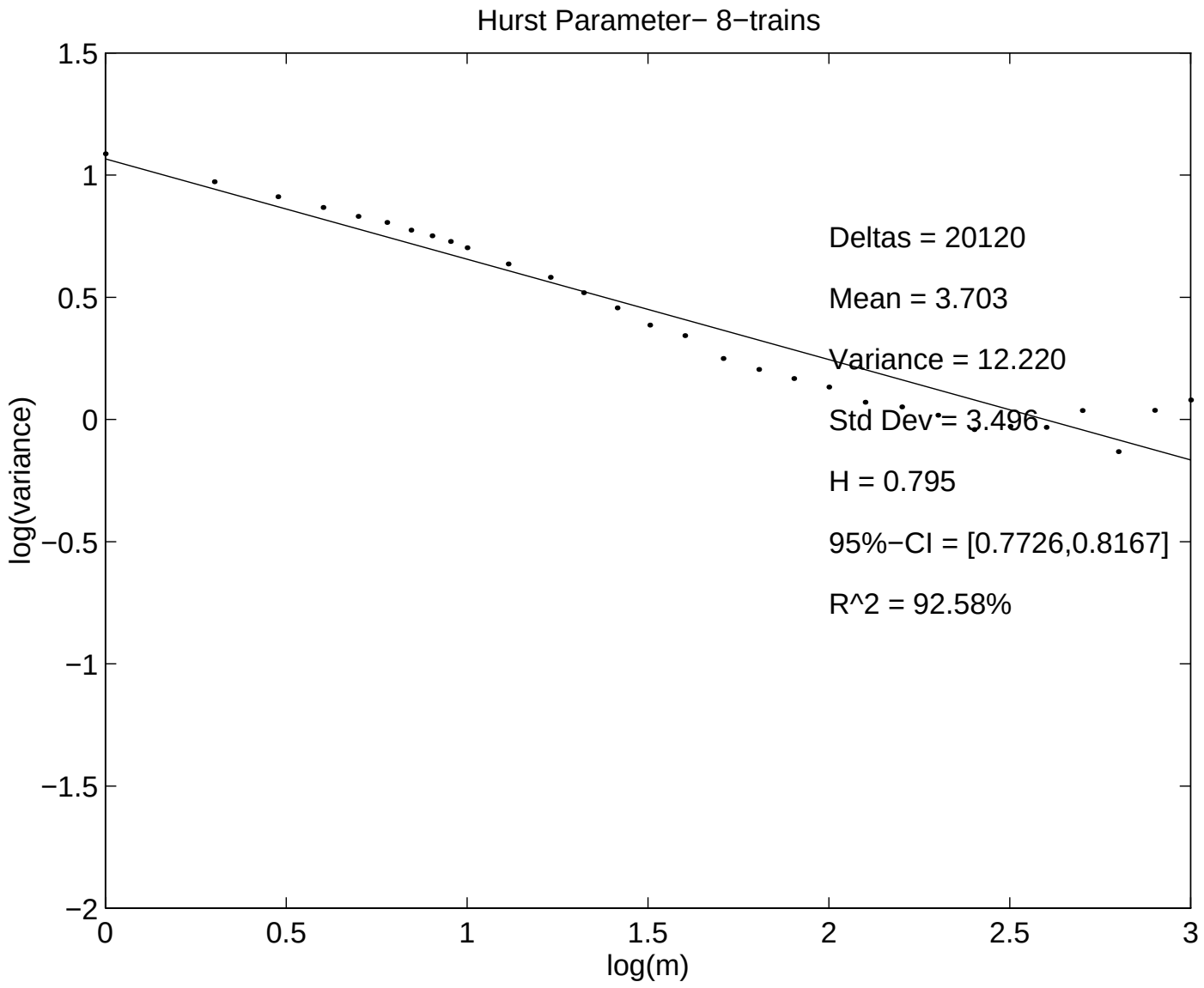

(b) 8 Generated Packet Trains

Figure 9: Graph 9f & 9g

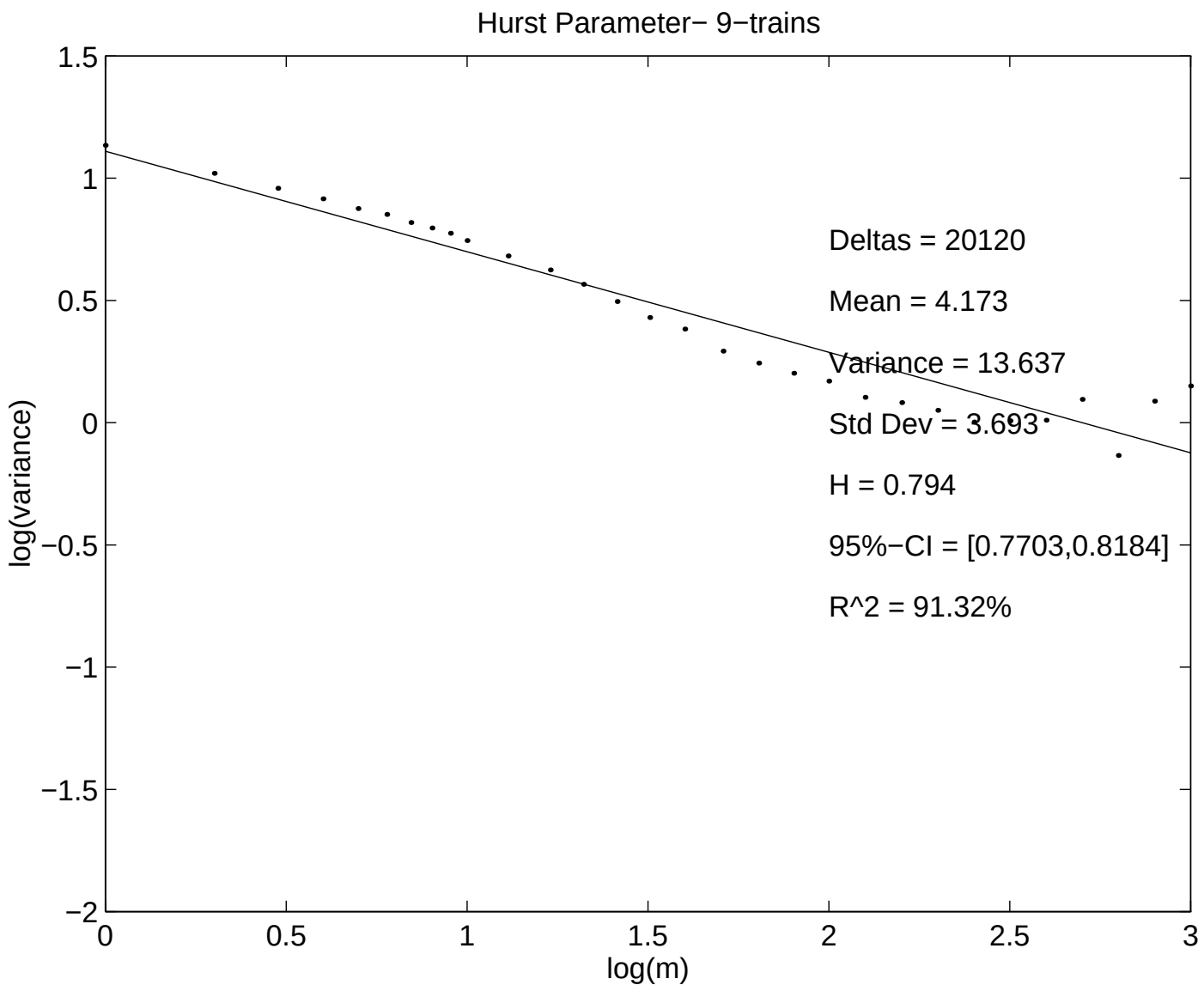

(a) 9 Generated Packet Trains

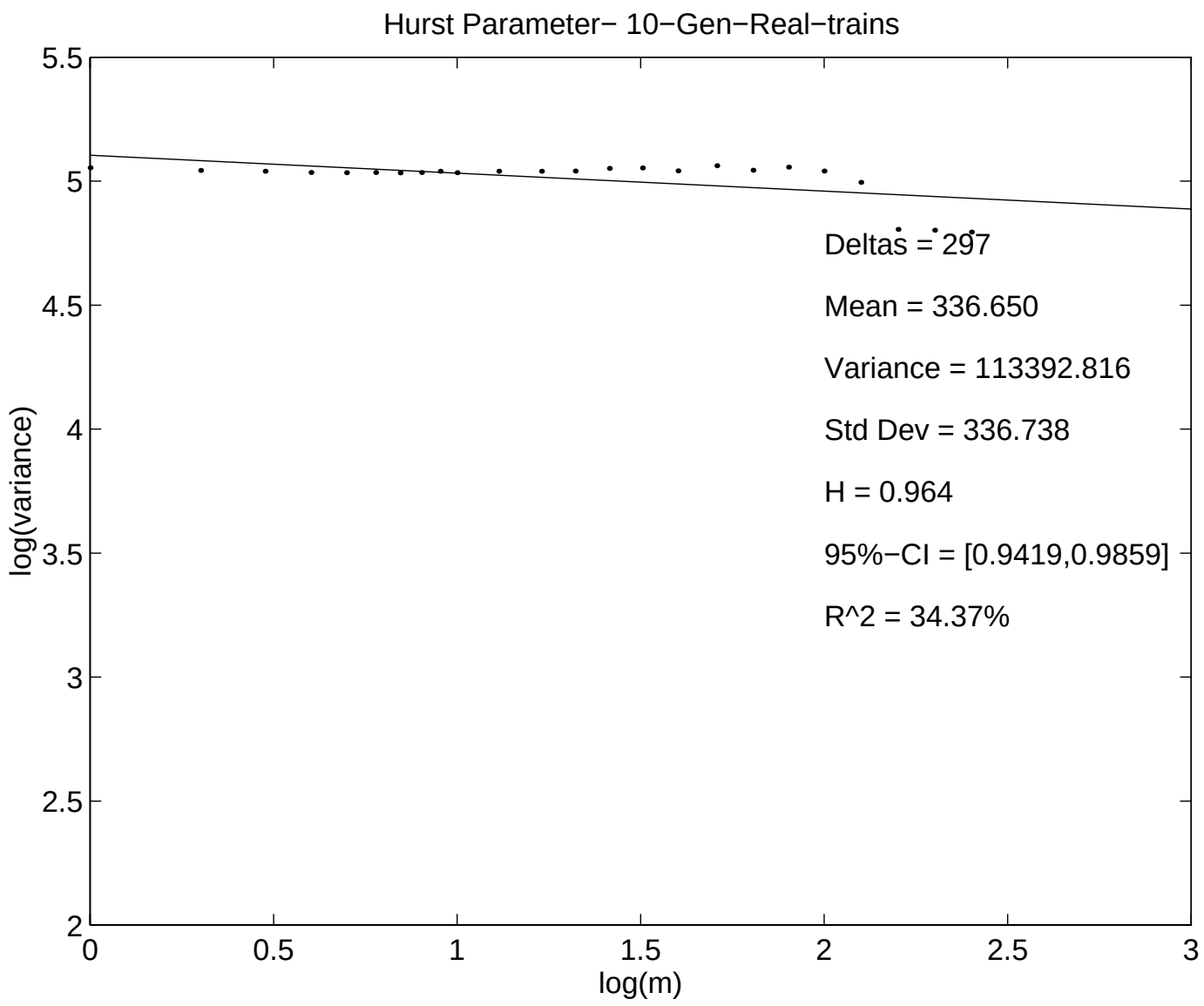

(b) 10 Generated Real Packet Trains

Figure 10: Graph 9h & 10

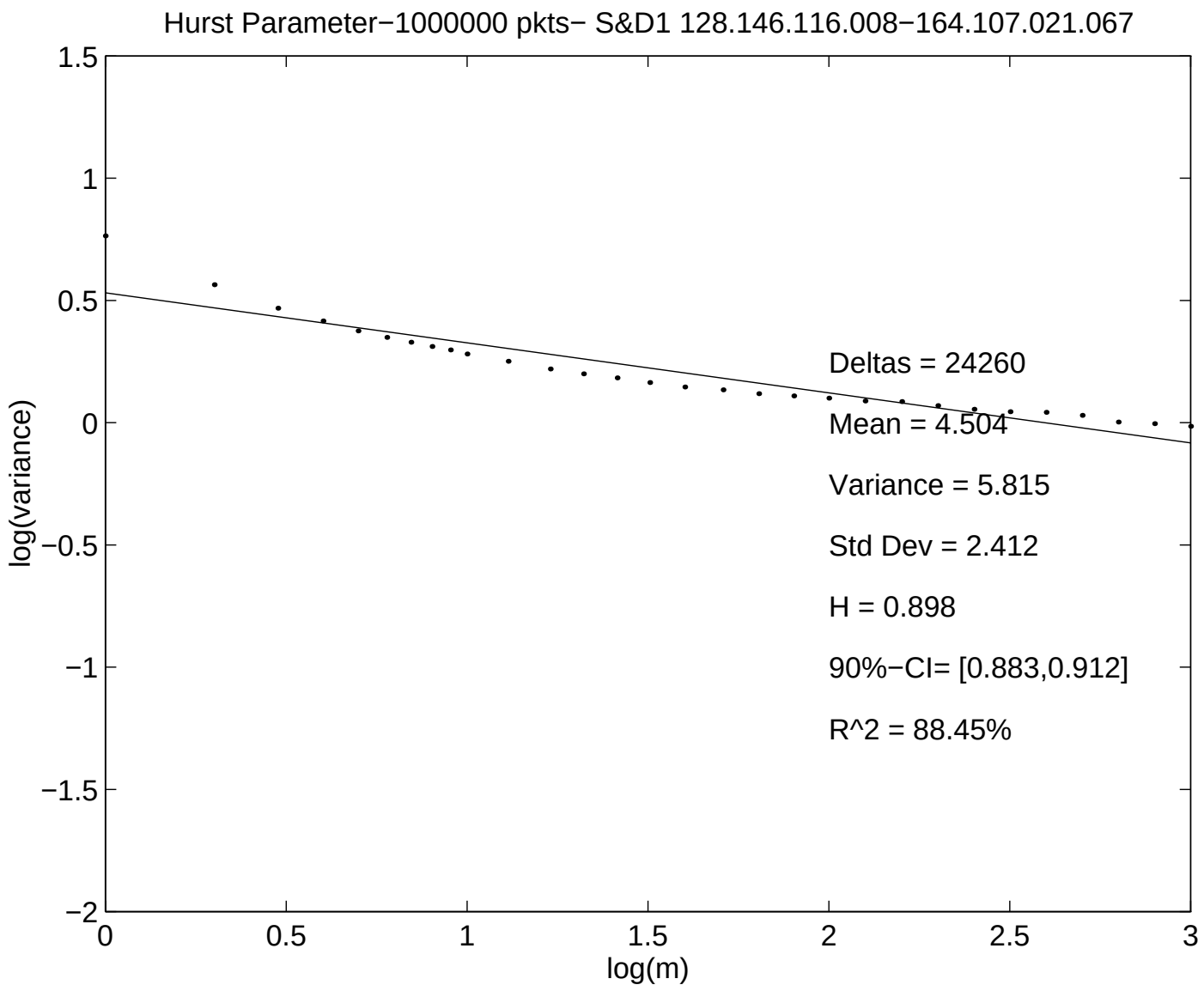

(a) Both Src & Dest 1

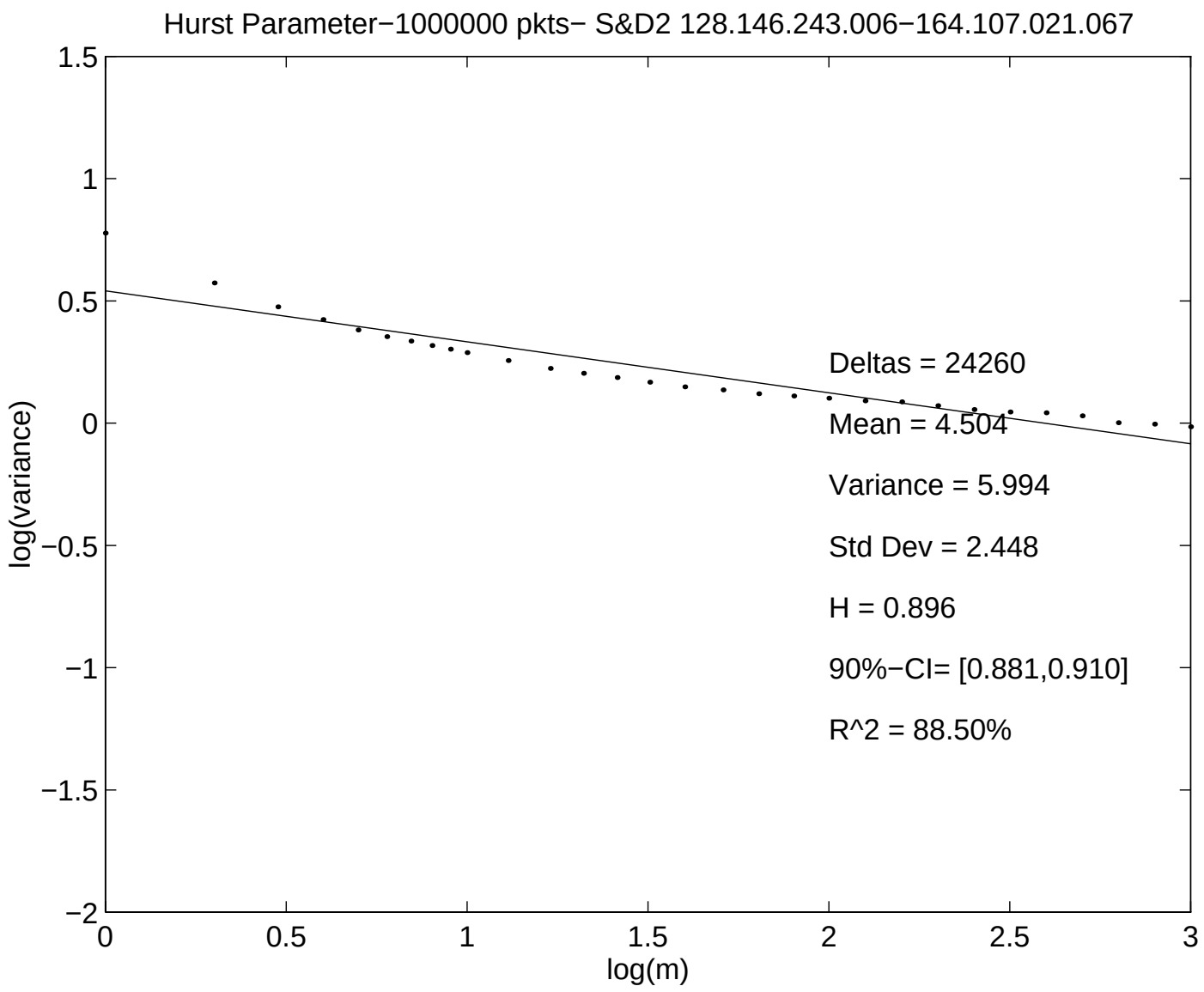

(b) Both Src & Dest 2

Figure 11: Graph 11a & 11b

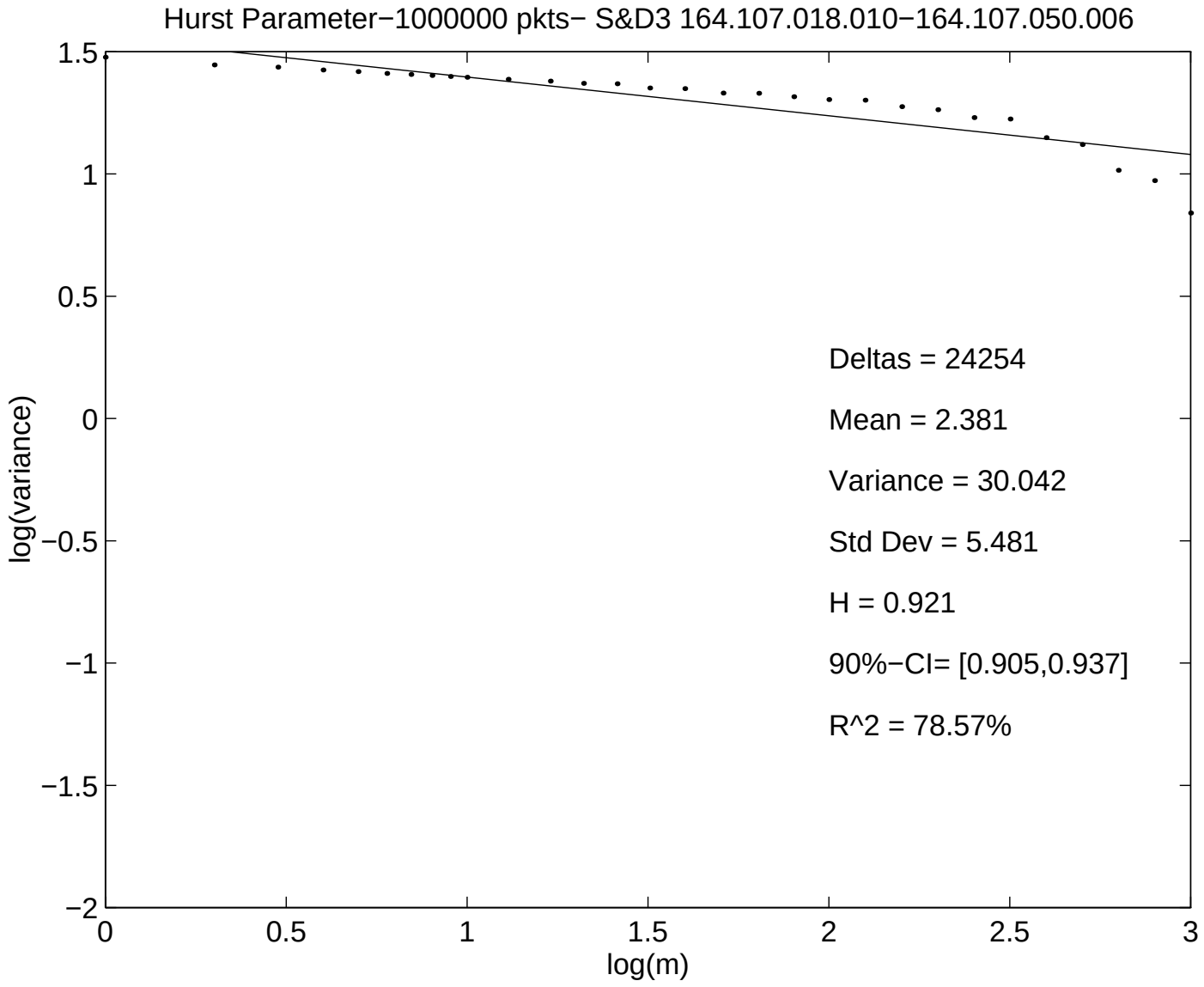

(a) Both Src & Dest 3

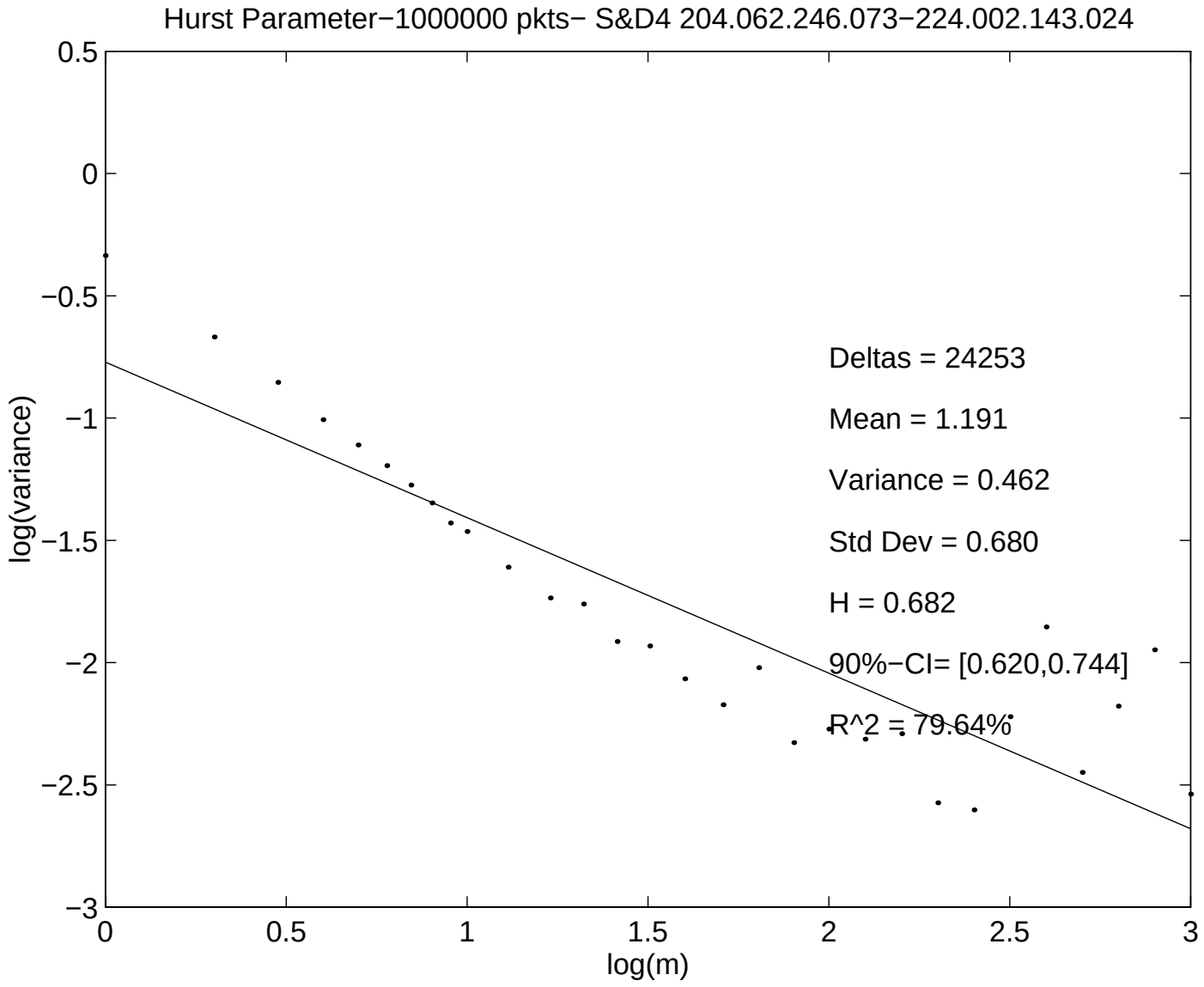

(b) Both Src & Dest 4

Figure 12: Graph 11c & 11d

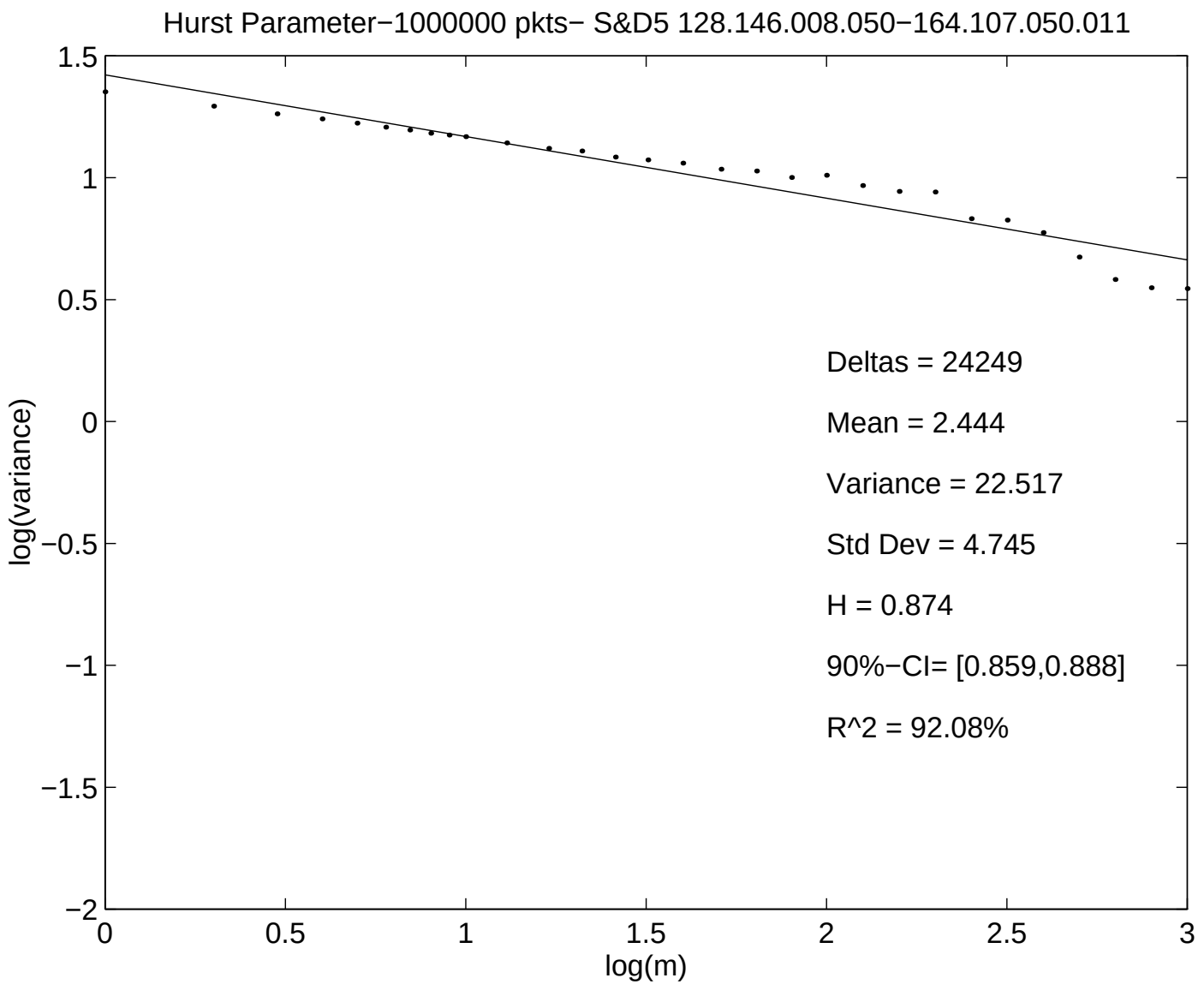

(a) Both Src & Dest 5

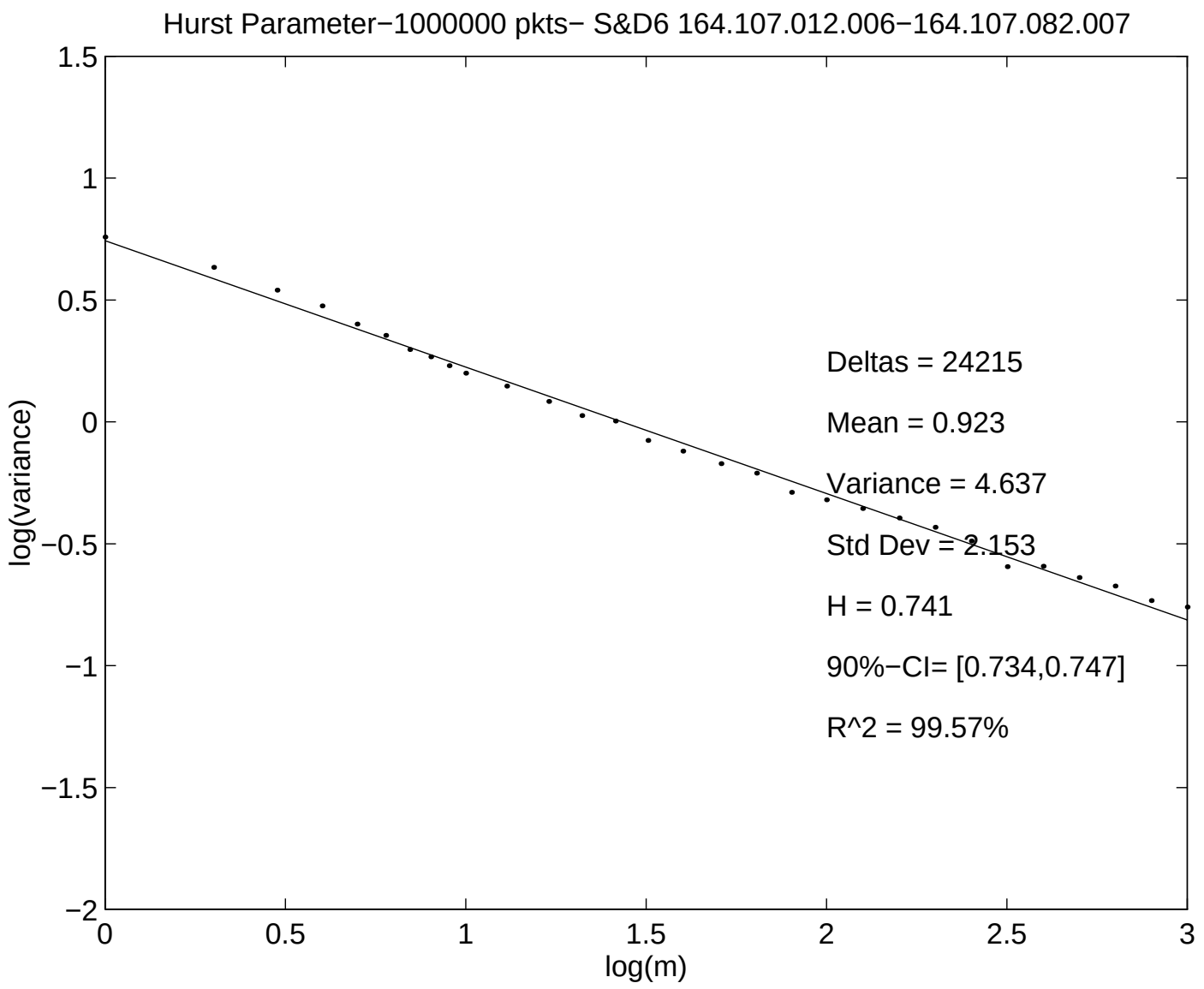

(b) Both Src & Dest 6

Figure 13: Graph 11e & 11f

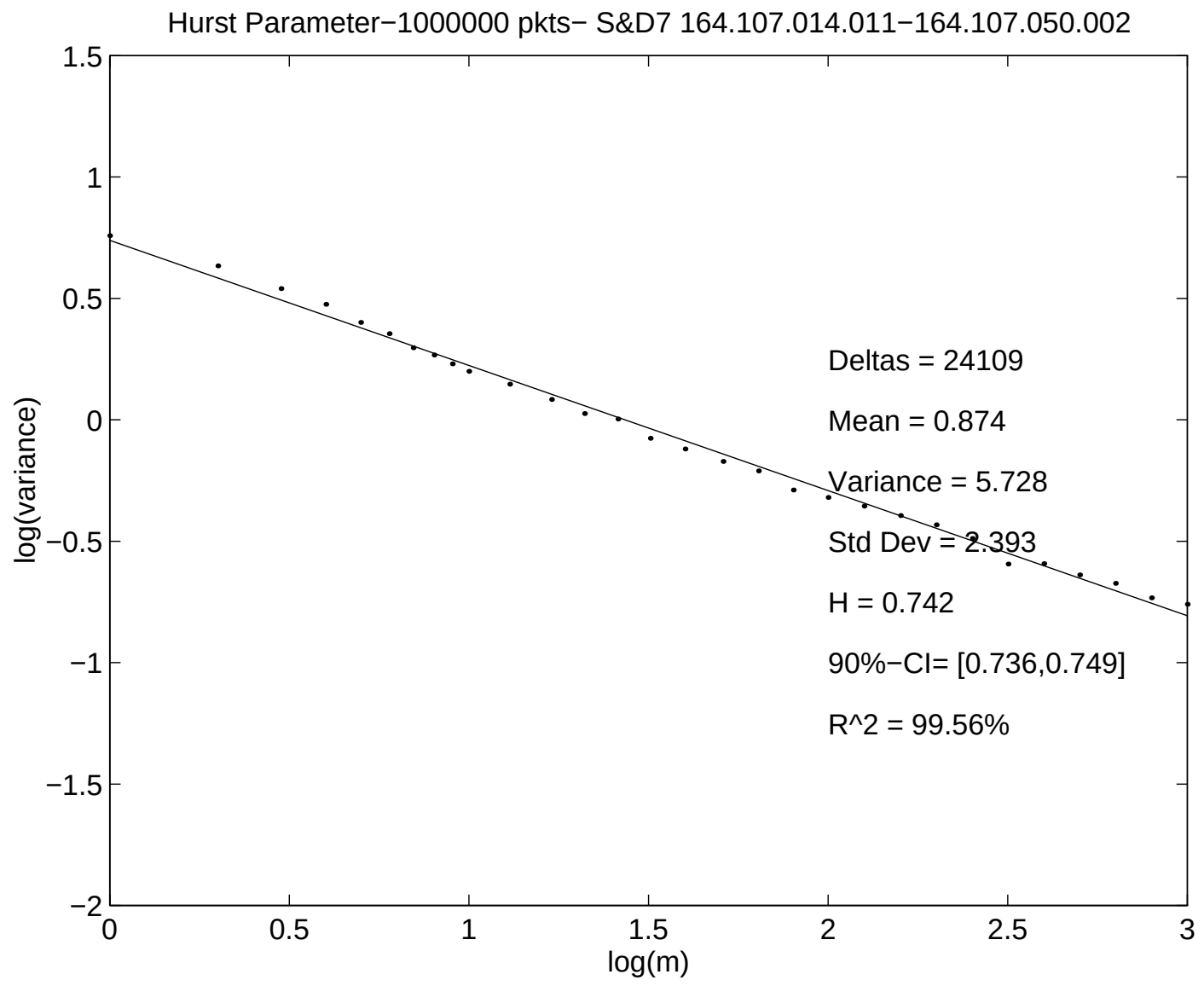

(a) Both Src & Dest 7

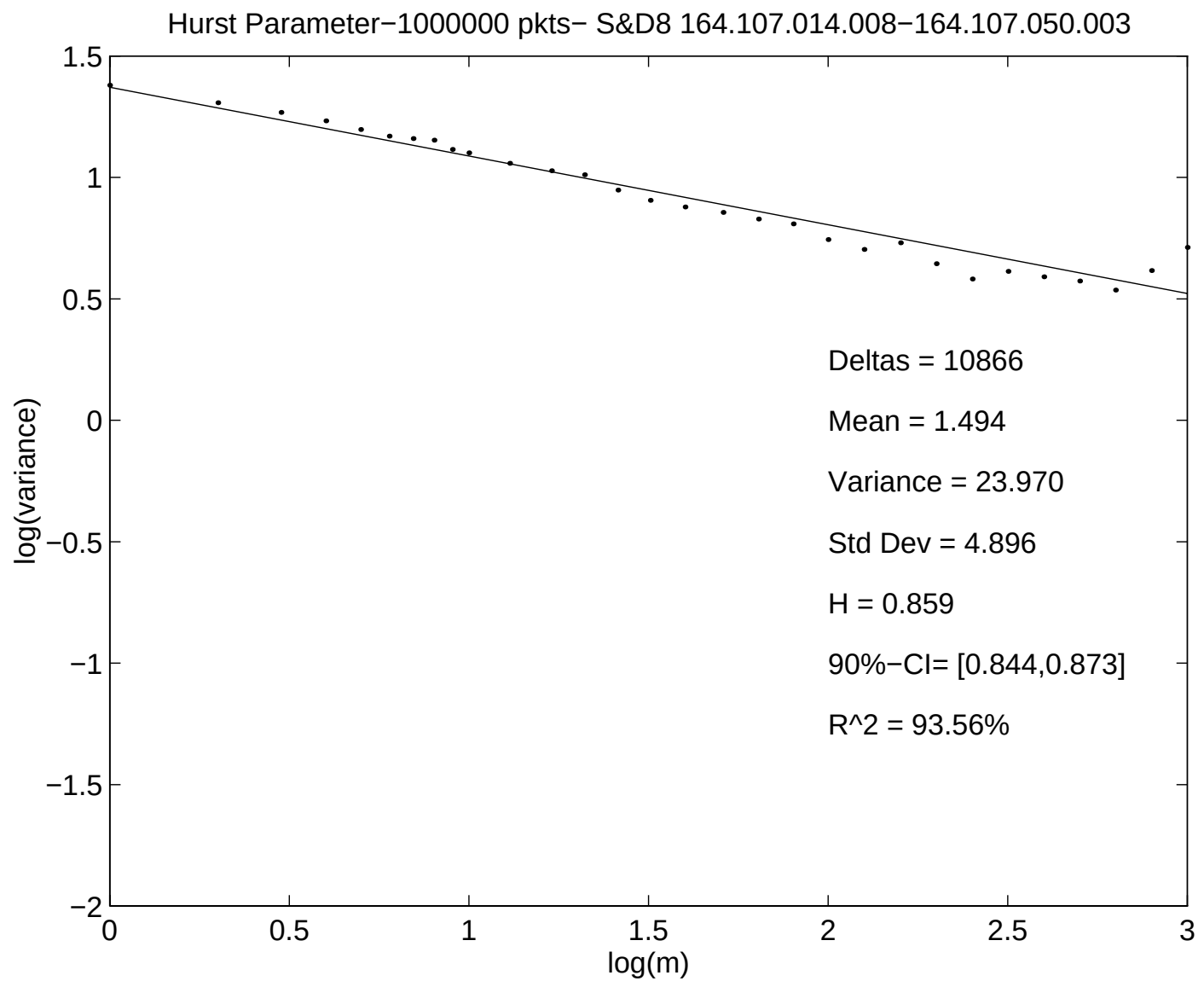

(b) Both Src & Dest 8

Figure 14: Graph 11g & 11h

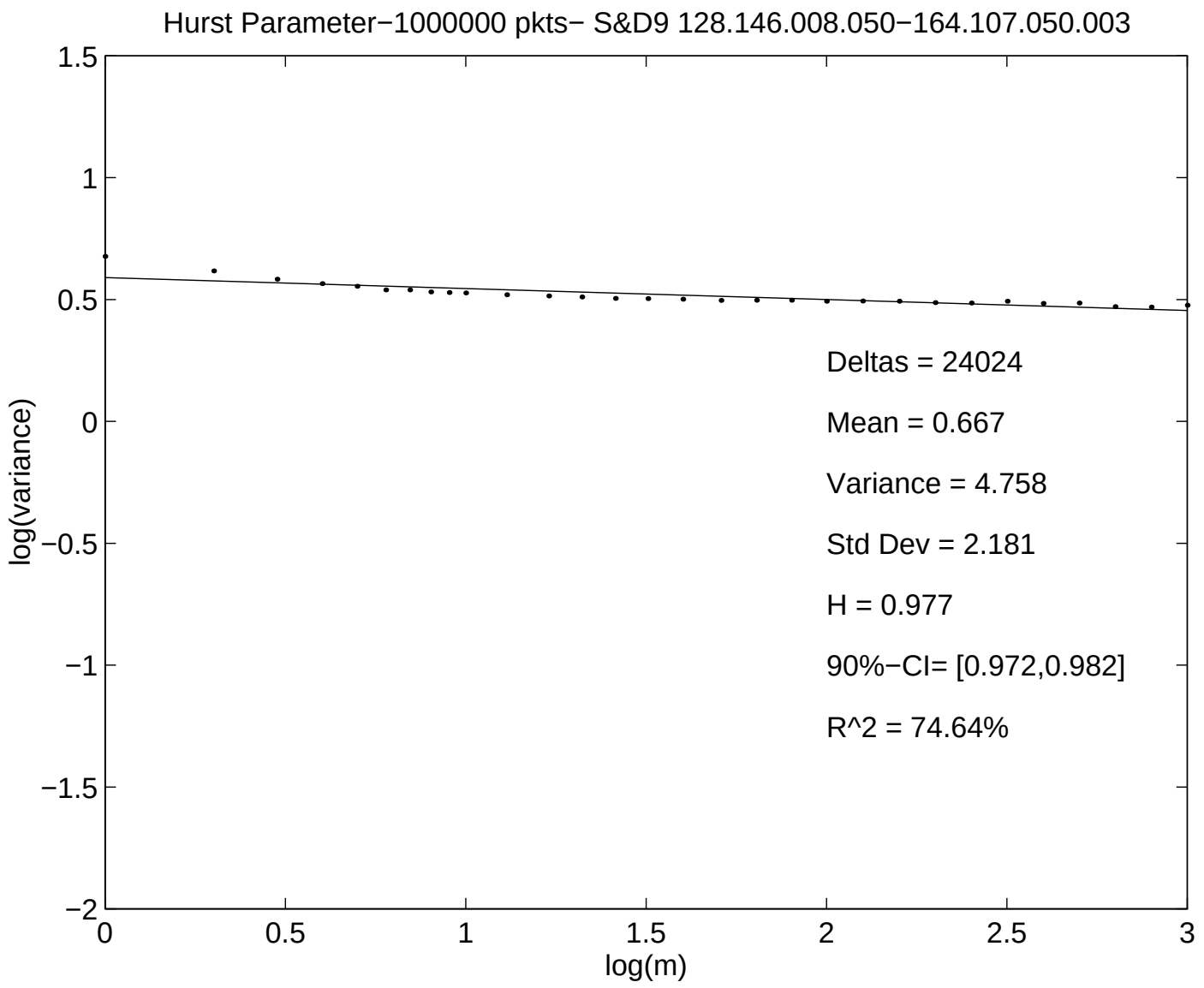

(a) Both Src & Dest 8

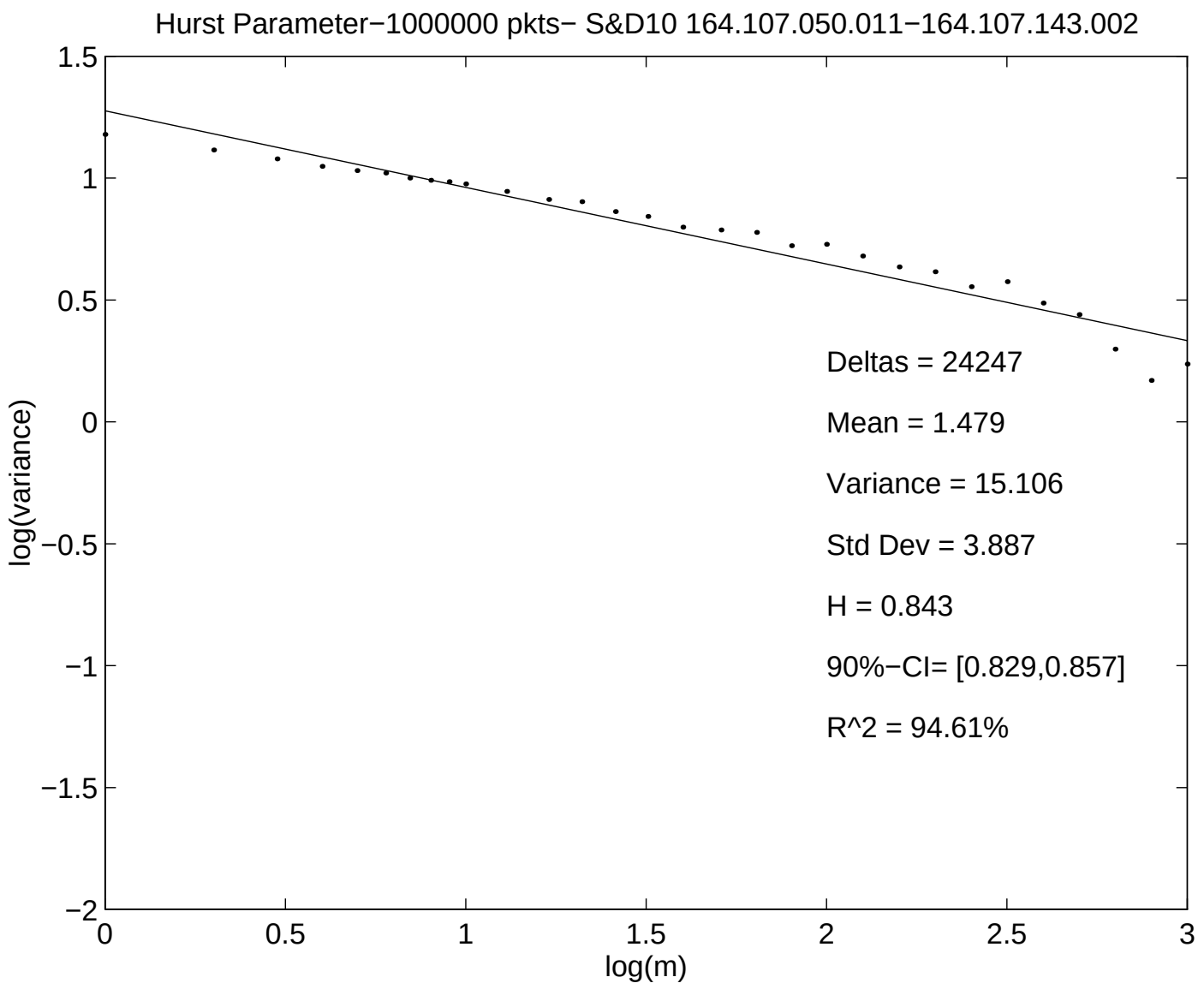

(b) Both Src & Dest 10

Figure 15: Graph 11i & 11j